\documentclass{article}
\usepackage[margin=1in]{geometry}
\usepackage[utf8]{inputenc}
\usepackage{amssymb}
\usepackage{bbm}	
\usepackage{amsthm,amssymb,amsmath}
\usepackage{mathtools}
\usepackage{thmtools,thm-restate}
\usepackage{tikz}
\usetikzlibrary{arrows,automata}
\usepackage{multirow}
\usepackage{tabularx}
\usepackage[ruled,lined,boxed]{algorithm2e}
\usepackage{algorithmic}
\usepackage{comment}
\usepackage[caption=false]{subfig}
\usepackage{booktabs}
\usepackage{libertine}
\usepackage{authblk}
\usepackage[libertine]{newtxmath}
\usepackage{cuted}
\usepackage{flushend}
\usepackage{subfiles}
\newtheorem{obs}{Observation}
\newtheorem{remark}{Remark}
\newtheorem{exmp}{Example}
\newtheorem{axiom}{Property}

\usepackage{thm-restate}
\usepackage[libertine]{newtxmath} % must load after ams packages
\usepackage[scaled=0.96]{zi4} % use Inconsolata as the typewriter font

\usepackage{a4wide, dsfont, microtype, xcolor, paralist}

\usepackage[ocgcolorlinks]{hyperref} % Option ocgcolorlinks makes links only colored on screen and not on printouts
\colorlet{DarkRed}{red!50!black}
\colorlet{DarkGreen}{green!50!black}
\colorlet{DarkBlue}{blue!50!black}
\hypersetup{
	linkcolor = DarkRed,
	citecolor = DarkGreen,
	urlcolor = DarkBlue,
	bookmarksnumbered = true,
	linktocpage = true
}

\newcommand{\HG}[1]{{\small\textcolor{magenta}{HG: #1}}}

\let\epsilon\varepsilon

\usepackage{xspace}

\usepackage{tikz}
\usetikzlibrary{decorations.pathreplacing}
\usetikzlibrary{plotmarks}
\usetikzlibrary{positioning,automata,arrows}
\usetikzlibrary{shapes.geometric}
\usetikzlibrary{decorations.markings}
\usetikzlibrary{positioning, shapes, arrows}

\PassOptionsToPackage{usenames,dvipsnames,svgnames}{xcolor}
\definecolor{orange}{RGB}{235,90,0}
\definecolor{darkorange}{RGB}{175,30,0}
\definecolor{turkis}{RGB}{131,182,182}
\definecolor{darkturkis}{RGB}{31,82,82}
\definecolor{green}{RGB}{102,180,0}
\definecolor{darkgreen}{RGB}{51,90,0}
\definecolor{myblue}{RGB}{0,0,213}
\definecolor{mydarkblue}{RGB}{0,0,100}
\definecolor{mybrightblue}{HTML}{74B0E4}
\definecolor{mybrighterblue}{HTML}{B3EAFA}
\definecolor{lila}{RGB}{102,0,102}
\definecolor{darkred}{RGB}{139,0,0}
\definecolor{darkyellow}{RGB}{188,135,2}
\definecolor{brightgray}{RGB}{200,200,200}
\definecolor{darkgray}{RGB}{50,50,50}
\definecolor{amaranth}{rgb}{0.9, 0.17, 0.31}
\definecolor{alizarin}{rgb}{0.82, 0.1, 0.26}
\definecolor{amber}{rgb}{1.0, 0.75, 0.0}
\definecolor{green(ryb)}{rgb}{0.4, 0.69, 0.2}
\definecolor{hanblue}{rgb}{0.27, 0.42, 0.81}
\definecolor{grannysmithapple}{rgb}{0.66, 0.89, 0.63}

\newtheorem{theorem}{Theorem}[section]
\newtheorem{lemma}[theorem]{Lemma}
\newtheorem{proposition}[theorem]{Proposition}
\newtheorem{definition}[theorem]{Definition}

\newtheorem{fact}[theorem]{Fact}
\newtheorem{lemma-rstbl}[theorem]{Lemma}
\newtheorem{obs-rstbl}[theorem]{Observation}
\newtheorem{theorem-rstbl}[theorem]{Theorem}

\usepackage{environ}
\NewEnviron{cproblem}[1]{%
\begin{center}\fbox{\parbox{5.5in}{%
    {\centering\scshape #1\par}%
    \parskip=1ex
    \everypar{\hangindent=1em}%
    \BODY
}}\end{center}}

\title{Computation and Bribery of Voting Power in Delegative Simple Games}
\author[1]{Gianlorenzo D'Angelo}
\author[1]{Esmaeil Delfaraz}
\author[2]{Hugo Gilbert}
\affil[1]{\normalsize Gran Sasso Science Institute, L'Aquila, Italy}
\affil[2]{Université Paris-Dauphine, Université PSL, CNRS, LAMSADE, 75016 Paris, France}
\date{}

\begin{document}

\maketitle

\begin{abstract}
%Weighted voting games is one of the most important classes of cooperative games. 
    %Recently, Zhang and Grossi~\cite{zhang2020power} proposed a variant of this class, called \textit{delegative simple games}, which is well suited to analyse the relative importance of each voter in liquid democracy elections. Moreover, they defined a power index, called the delagative Banzhaf index to compute the importance of each agent (i.e., both voters and delegators) in a delegation graph based on two key parameters: the total voting weight she has accumulated and the structure of supports she receives from her delegators. 
    
    Following Zhang and Grossi~(AAAI 2021), we study in more depth a variant of weighted voting games in which agents' weights are induced by a transitive support structure. 
    This class of simple games is notably well suited to study the relative importance of agents in the liquid democracy framework. %Banzhaf and Shapley-Shubik indices in liquid democracy. 
    We first propose a pseudo-polynomial time algorithm to compute the Banzhaf and Shapley-Shubik indices for this class of game. %delegative Banzhaf and Shapley-Shubik values in delegative simple games.
    Then, we study a bribery problem, in which one tries to maximize/minimize the voting power/weight of a given agent by changing the support structure under a budget constraint. 
    We show that these problems are computationally hard and provide several parameterized complexity results. %Lastly, we investigate a power distribution problem in which one tries to find a delegation graph with a predefined number of gurus to maximize the minimum power index value of an agent. We show that this is an NP-hard problem even when the weight of each voter is $1$.
\end{abstract}

\section{Introduction}\label{secIN}
Weighted Voting Games (WVG) form a simple scheme to model situations in which voters must make a yes/no decision about accepting a given proposal~\cite{chalkiadakis2016weighted}. 
Each voter has a corresponding weight and the proposal is accepted if the sum of weights of agents supporting the proposal exceeds a fixed threshold called the quota. 
In a WVG, weights of voters can represent an amount of resource and the quota represents the quantity of this resource which should be gathered to enforce the proposal. 
For instance, agents may be political parties and weights could be derived from the relative importance of each party in terms of number of votes received. 
There is a large literature to measure the relative importance of each agent in such a situation~\cite{chalkiadakis2016weighted,felsenthal1998measurement}. 
Notably, the computational and axiomatic properties of two such measures, the Banzhaf measure of voting power and the Shapley-Shubik index~\cite{banzhaf1964weighted,shapley1953value}, have been extensively studied.  
The computational properties investigated obviously include the computational complexity of computing these measures for a given agent~\cite{deng1994complexity} but they also involve several manipulation problems~\cite{aziz2011false,felsenthal1998measurement,zick2011,zuckerman2012manipulating}.  

One possible limitation of WVGs is that they consider each agent as an indivisible entity and may not be able to represent agents who are composed of a complex structure. However, if agents are corporations or political parties, then they do have some inner structure which may have an impact on their relative strength. 

\begin{figure}
    \centering
\scalebox{1}{
\begin{tikzpicture}
				\tikzset{vertex/.style = {shape=circle,draw = black,thick,fill = white}}
				\tikzset{edge/.style = {->,> = latex'}}
				
				\node[] (Pi) at  (-2,0) {$\Pi$};
				
				\node[vertex] (A) at  (0,0) {A};
				\node[] (Aw) at  (0.5,0) {3};
				\node[vertex] (B) at  (-1,-1) {B};
				\node[] (Bw) at  (-0.5,-1) {2};
				\node[vertex] (C) at  (1 ,-1) {C};
				\node[] (Cw) at  (1.5,-1) {4};
				\node[vertex] (D) at  (-2,-2) {D};
				\node[] (Dw) at  (-1.5,-2) {1};
				\node[vertex] (E) at  (0,-2) {E};
				\node[] (Ew) at  (0.5,-2) {2};
				
				\draw[] (-3,0.6) to (2,0.6) to (2,-2.6) to (-3,-2.6) to (-3,0.6);

				%edges
				\draw[edge] (B) to (A);
				\draw[edge] (C) to (A);
				\draw[edge] (D) to (B);
				\draw[edge] (E) to (B);
				
				\end{tikzpicture}}
\scalebox{1}{
\begin{tikzpicture}
				\tikzset{vertex/.style = {shape=circle,draw = black,thick,fill = white}}
				\tikzset{edge/.style = {->,> = latex'}}
				
				\node[] (Pip) at  (4,0) {$\Pi'$};
				
				\node[vertex] (A') at  (5.5,0) {A'};
				\node[] (A'w) at  (6,0) {3};
				\node[vertex] (B') at  (4 , -1.5) {B'};
				\node[] (B'w) at  (4.5 , -1.5) {3};
				\node[vertex] (C') at  (5.5 ,-1.5) {C'};
				\node[] (C'w) at  (6,-1.5) {2};
				\node[vertex] (D') at  (7,-1.5) {D'};
				\node[] (D'w) at  (7.5,-1.5) {3};

				\draw[] (3,1.1) to (8,1.1) to (8,-2.1) to (3,-2.1) to (3,1.1);

				%edges
				\draw[edge] (B') to (A');
				\draw[edge] (C') to (A');
				\draw[edge] (D') to (A');
		\end{tikzpicture}}
    \caption{Two political parties $\Pi$ and $\Pi'$ with two different inner structures.}
    \label{fig: parties pi and pip}
\end{figure}
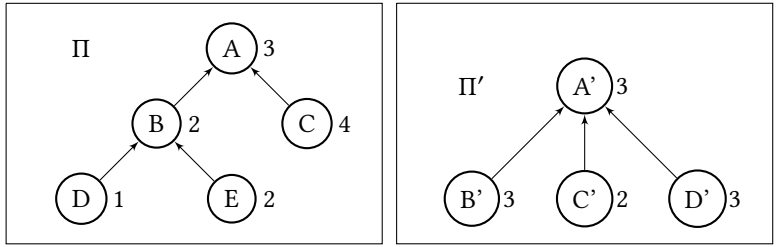

\begin{exmp}\label{example: two political parties}
Let us consider two political parties $\Pi$ and $\Pi'$ represented in Figure~\ref{fig: parties pi and pip}. 
Each party has a political leader (agents $A$ and $A'$ respectively) and different inner political trends with subleaders (agents $B$, $C$, $D$, $E$ in $\Pi$ and agents $B'$, $C'$, and $D'$ in $\Pi'$). 
Each leader and subleader have their own supporters which provide them some voting weights, written next to each agent in Figure~\ref{fig: parties pi and pip}. 
In each party, the different agents form a directed tree structure of support, where each arc represents the fact that an agent supports another agent.  
For instance, in party $\Pi$, agent $D$ endorses agent $B$ which herself endorses agent $A$. 
In this way, agent $D$ implicitly supports agent $A$ and puts at her disposal her voting weight.  
Note however that it may not be possible for agent $D$ to directly endorse agent $A$. Indeed, if agent $D$ represents the most left-wing sensibility of the party whereas agent $A$ represents a more consensual political trend, it may be difficult for agent $D$ to publicly support agent $A$ without losing credibility in the eye of her supporters.  
By delegating to $A$ through $B$, agent $D$ indicates that her support to $A$ is conditioned to the presence of agent $B$.

The total weight accumulated by voters $A$ and $A'$ are respectively worth 12 and 11. Hence, the total weight gathered by $A$ is greater than the one of $A'$. Could this indicate that agent $A$ is at least as powerful as agent $A'$. The inner structures of parties $\Pi$ and $\Pi'$ suggest otherwise. Indeed, agent $A$ receives a greater total weight and is endorsed (directly and indirectly) by more agents. However, agent $A'$ receives direct support from all other subleaders of her party  which is not the case of agent $A$. As a result, if agent $B$ decides to secede and create her own party, then agent $A$ would lose the support of agents $D$ and $E$ conceding a total weight lose of 5. Conversely, the most important weight lose that agent $A'$ can suffer from the secession of another agent is worth 3. Hence, the inner structure of party $\Pi'$ which supports candidate $A'$ seems more robust than the one of candidate $A$. 
\end{exmp}

Example~\ref{example: two political parties} suggests to study a more complex model than the one of WVGs where agents are backed up by an internal support structure with transitive supports. Studying this kind of model has been recently initiated by Zhang and Grossi~\cite{zhang2020power}. 
Indeed, the authors investigate how to measure the relative importance of voters in the framework of \emph{Liquid Democracy} (LD)~\cite{behrens2014principles,ford2020liquid}. 
LD is a collective decision paradigm in which agents can vote themselves or delegate their vote to another agent. 
One important feature, is that an agent who receives delegations can in turn delegate her vote and the ones that she has received to another agent which is exactly the kind of transitive support discussed in Example~\ref{example: two political parties}. 
In their work, Zhang and Grossi~\cite{zhang2020power} define \emph{Delegative Simple Games} (DSG), a variant of WVGs in which the capacity of a subset of agents to reach the quota does not only depend on their voting weights but also on their delegations. 
DSGs make it possible to take into account the support (delegation) structure underlying the game and to favor agents who receive more direct supports, compared to agents receiving more distant chains of support.  
The authors notably study axiomatically the Banzhaf measure of voting power applied to this kind of game which they term the \emph{delegative Banzhaf measure of voting power}.\footnote{This measure was in fact termed, the delegative Banzhaf index. However, following Felsenthal and Machover~\cite{felsenthal1998measurement}, we reserve the term index for measures whose values sum up to one when considering all agents.} 

\paragraph{Our contribution.} 
We obtain several results related to DSGs.
We first study several properties, notably computational properties, of the delegative Banzhaf and Shapley-Shubik measures of power.  
For instance, while the computation of these measures is computationally hard, we show that they can be calculated by a pseudo-polynomial dynamic programming algorithm similar to the one for Banzhaf and Shapley-Shubik measures in WVGs. 
We then investigate a bribery problem where, given a delegation graph, the goal is to maximize/minimize the voting power/weight of an agent by changing at most a fixed number of delegations. 
We show that the problems related to bribing voting power are hard, and that the maximization problems are hard to approximate even when the social network is a tree. 
We then move to the conceptually simple bribery problems related to voting weight. 
On these problems, we obtain both hardness and tractability results by investigating the approximation and the parameterized complexity viewpoints. 
%We show that the problems of minimizing/maximizing a voter's voting power are strongly NP-hard. Furthermore, we prove that having a better approximation guarantee than $1-1/e$ to maximize the voting weight of a voter is not possible, unless $P = NP$, then we provide some parameterized complexity results for this problem. 

    %Finally, we show that finding a delegation graph with a given number of gurus that maximizes the minimum power index value an agent can have is a computationally hard problem.
    %In this work, we study in more depth the delegative Banzhaf and Shapley-Shubik indices. While the computation of these indices is computationally hard, we show that they can be calculated by a pseudo-polynomial dynamic programming algorithm similar to the one of standard Banzhaf and Shapley-Shubik indices. %Second, we provide an axiomatic characterization of the delegative Shapley-Shubik index.
%Second, we study a bribery problem, in which one tries to maximize/minimize the voting power/weight of a given agent by changing at most $k$ delegation choices. We show that these problems are NP-hard and provide several parameterized complexity results. %Lastly, we investigate a power distribution problem in which one tries to find a delegation graph with a predefined number of gurus to maximize the minimum power index value of an agent. We show that this is an NP-hard problem even when the weight of each voter is $1$.
\section{Related Work}\label{secRW}
WVGs originated in the domain of cooperative game theory~\cite{chalkiadakis2011computational,chalkiadakis2016weighted} and are used to study the a-priori voting power of voters in an election~\cite{felsenthal1998measurement}. 
Two well-known solutions to measure the importance of an agent in a WVG are the Shapley-Shubik index and the Banzhaf measure of voting power~\cite{banzhaf1964weighted,shapley1953value}. 
%Both indices share several properties~\cite{dubey1979mathematical}. For instance, they both satisfy the well-known axioms \textit{Dummy Player}, \textit{Symmetry} and \textit{Additivity}. Several works characterize both indices in different ways which emphasize their differences~\cite{barua2005new,van2002axiomatization,neyman1989uniqueness,shapley1953value}.
Computing these measures is \#P-Complete~\cite{deng1994complexity,prasad1990np}. 
However, Matsui and Matsui~\cite{matsui2000survey} designed pseudopolynomial algorithms that can compute the Shapley-Shubik and Banzhaf measures in time $O(n^3w_{\max})$ and $O(n^{2}w_{\max})$ respectively, where $n$ is the number of agents and $w_{\max}$ is the maximum weight of an agent. 
Other works have been dedicated to their computation, either to compute them exactly~\cite{brams1976power,conitzer2004computing,ieong2005marginal,ieong2006multi}, or approximately~\cite{bachrach2008approximating,fatima2010approximation,leech2003computing,mann1960values,merrill1982approximations}. 
Moreover, several manipulation problems involving voting power measures have been investigated, as computing the quota maximizing or minimizing the importance of a specific voter~\cite{zick2011,zuckerman2012manipulating} or determining the impact of splitting an agent in two~\cite{aziz2011false,felsenthal1998measurement,laruelle2005critical}. %fatima2008linear, owen1988multilinear

Several works have studied the measurement of voters' importance in an LD election~\cite{boldi2009voting,boldi2011viscous,kling2015voting,zhang2020power}.  
Boldi et al.~\cite{boldi2009voting,boldi2011viscous} proposed a way to measure the relative importance of each voter in a social network using a power index similar to PageRank. 
In their model, each voter can recommend one of her neighbor in the network. 
The recommendations of the voters are then transitively delegated but are attenuated by using a multiplicative damping factor. 
The authors term the resulting system a viscous democracy election as the voting power does not flow completely but rather meets some resistance. %due to the damping factor. 
In this way, the model by Boldi et al.~\cite{boldi2009voting,boldi2011viscous} favors the voters who receive direct supports instead of more distant ones.
Kling et al.~\cite{kling2015voting} studied the behavior of voters using the LiquidFeedback software in the German Pirate Party. The authors studied the number and types of interactions with the software, the distribution in terms of number of delegations received per voter as well as the behavior of ``super-voters'' which receive a large number of delegations. 
The authors then applied several power measures (e.g., Shapley-Shubik and Banzhaf) to analyze the power distribution in LD elections. 
This analysis led the authors to propose modifications to the existing power measures to fit better to the data they gathered. 
More precisely, the authors designed generalizations of theses measures which allow to model non-uniform distributions of approval rates. %\footnote{Stated otherwise, the assumption that voters vote with equal probability in favour or against a proposal is not made.}. 
Lastly, as detailed in the introduction, Zhang and Grossi~\cite{zhang2020power} have recently proposed a formal way to measure the influence of voters in an LD election by introducing a variant of WVGs. This variant as well as the resulting power measures will be presented in Section~\ref{sec : Preliminaries}.

\section{Preliminaries} \label{sec : Preliminaries}
%In this section, we formalize delegative simple games and specify our setting.

%For each $i \in \mathbb{N}$, we write $[i]=\{1, \dots, i\}$. 
\subsection{Weighted voting games} % and power measures}
A simple game is a tuple $\mathcal{G} = \langle V, \nu \rangle$, where $V = [n]$ is a set of $n$ agents and $\nu : 2^n\rightarrow\{0,1\}$ is a characteristic function which only takes values 0 and 1. The notation $[i]$ and $[i]_0$ will denote the sets $\{1,\ldots i\}$ and  $\{0,1,\ldots, i\}$ respectively.  
A subset $C\subseteq V$ will also be called a coalition. 
For any coalition $C \subseteq V$, $C$ is said to be a winning (resp. losing) coalition if $\nu(C) = 1$ (resp. 0). % otherwise, it is said to be a losing coalition. 
An agent $i$ is said to be a swing agent for coalition $C$ if $\delta_i(C):=\nu(C\cup\{i\}) - \nu(C)$ equals 1. 

%Weighted voting games are particular simple games which admit a compact representation.
In WVGs, there exists a quota $q$ and each agent (also called voter) $i$ is associated with a weight $w_i$. The characteristic function $\nu$ is then defined by $\nu(C) = 1$ iff $\sum_{i \in C} w_i \ge q$. 
Stated otherwise, a coalition is winning if the sum of weights of agents in the coalition exceeds the quota. 
Several ways of measuring the importance of an agent in WVGs have been studied. 
We recall two of the most well known: 

\begin{definition}
The Banzhaf measure $B_i(\mathcal{G})$ and Shapley-Shubik index $Sh_i(\mathcal{G})$ of a voter $i$ in a simple game $\mathcal{G}$ are defined as 
$$B_i(\mathcal{G}) := \sum_{C\subseteq V\setminus\{i\}} \frac{1}{2^{n-1}} \delta_i(C),$$

$$Sh_i(\mathcal{G}) := \sum_{C\subseteq V\setminus\{i\}}\frac{1}{n}\frac{(n - |C| - 1)!|C|!}{(n-1)!}\delta_i(C).$$
\end{definition}

%\begin{definition}
%The Shapley-Shubik index of a voter $i$ in a simple game $\mathcal{G}$ is defined as 
%$$Sh_i(\mathcal{G}) := \sum_{C\subseteq V\setminus\{i\}}\frac{1}{n}\frac{(n - |C| - 1)!|C|!}{(n-1)!}\delta_i(C).$$
%\end{definition}
Hence, the Banzhaf measure and the Shapley-Shubik index provide two ways to measure the importance of a voter by quantifying her ability to be a swing agent. 
While both measures are worth investigation, they are quite different in nature as explained by Felsenthal and Machover~\cite{felsenthal1998measurement}. 
Indeed, while the Shapley-Shubik index is better explained as the expected share that an agent should earn from the election, seen as a game, the Banzhaf measure computes (based on a probabilistic model) the extent to which an agent is able to control the outcome of the election. 
Note that the first (resp. second) kind of measure is referred to as a notion of P-Power (resp. I-Power), where P stands for Prize (resp. I stands for Influence).  
As both measure have been extensively studied, we will study both in this paper.  
%The reader can refer to~\cite{felsenthal1998measurement} for details.
%In words, the Banzhaf and Shapley-Shubik indices of an agent $i$ measure how likely it is that $i$ is a swing agent for specific probability distributions on coalitions $C\subseteq V\setminus\{i\}$. For the Banzhaf index, each such coalition is equally likely. For the Shapley-Shubik index, the coalitions are drawn as follows: one first selects u.a.r. a number $s$ in $[n-1]_0$; then one selects a coalition $C\subseteq V\setminus\{i\}$ u.a.r. within coalitions of size $s$.

%In a weighted voting game, voters debate on a given proposal and their weights can be seen as amounts of resource related to the proposal at hand.  
%A coalition is then successful if it can gather enough of this resource to realize the proposal. For instance, voters may be heads of political parties and the weights could be derived from the relative importance of each party. However, as explained in the introduction, this simple intuition neglects the inner structure of each agent, which could be of importance.  Zhang and Grossi followed this intuition and defined delegative simple games, a variant of weighted voting games~\cite{zhang2020power} which makes it possible to take into account this structure. 
%Delegative simple games are motivated by liquid democracy elections.  

%\section{Simple Games and Liquid Democracy}
\subsection{A model of liquid democracy}
In the sequel, while our introduction suggests that DSGs can be used in a broader setting, we will follow the LD paradigm which showcases an interesting application where transitive support structures play a key role. This will notably be convenient to use the notations from Zhang and Grossi~\cite{zhang2020power}.\\

\textbf{A liquid democracy election.}
A finite set of agents $V = [n]$ will vote on a proposal. 
%Each agent is associated with a weight $w_i$, i.e., 
There is a weight function $\omega: V \longrightarrow \mathbb{N}_{>0}$, assigning a positive weight $\omega(i)=w_i$ to each voter $i$. 
A special case of interest is the one where all voters have weight one. 
The rule used is a super-majority rule with quota $q \in (\frac{\sum_{i \in V}w_i}{2},\sum_{i \in V}w_i] \cap \mathbb{N}_{>0}$.\footnote{In this work, we restrict the values of weights $w_i$ and $q$ to $\mathbb{N}_{>0}$ This restriction can be motivated by a result by Muroga~\cite{muroga1971threshold}.} Stated otherwise, the proposal is accepted if the total voting weight in favor of it is at least worth $q$. 

We assume the election to follow the LD paradigm. 
Notably, voters are vertices of a Social Network (SN) modeled as a directed graph $D = (V, A) $. Each node in the SN corresponds to a voter $i \in V$ and a directed edge $(i, j) \in A $ corresponds to a social relation between $i$ and $j$: it specifies that agent $i$ would accept to delegate her vote to $j$ or more generally speaking to endorse $j$.  The set of out-neighbors %(resp.\ in-neighbors) 
of voter $i$ is denoted by $\mathtt{Nb}_{out}(i)= \left\lbrace j \in V | (i,j) \in A \right\rbrace$.  
%(resp.\ $\mathtt{Nb}_{in}(i)= \left\lbrace j \in V | (j,i) \in A \right\rbrace$). 
Each agent $i$ has two possible choices: either she can vote directly, or she can delegate her vote to one of her neighbors in $\mathtt{Nb}_{out}(i)$. 
The information about delegation choices is formalized by a delegation function $d$, where
%\begin{itemize}
	%\item
	$ d(i)=j$ if voter $i$ delegates to voter $j \in \mathtt{Nb}_{out}(i)$,
	%\item
	and $d(i)=i$ if voter $i$ votes directly.

%\ED{We could maybe take into account cycles. In that case, we only need to change a few parts of the paper.} 
The delegation digraph $H_d = (V,E)$ resulting from $d$ is the subgraph of $D$, where $(i,j)\in E$ iff $i\neq j$ and $d(i) = j$. We assume that this digraph is acyclic. More precisely, we assume $H_d$ is a spanning forest of in-trees where all vertices have out-degree 1 except the roots which have out-degree 0. For any digraph $D$, $\Delta(D)$ denotes the set of all acyclic delegation graphs $H_d$ that can be induced by delegation functions $d$ on $D$.  In $H_d$, we denote by $c_d(i,j)$ the delegation chain which starts with voter $i$ and ends with voter $j$. Put another way, $c_d(i,j)$ is a sequence $((v_1,v_2),\ldots,(v_{k-1},v_k))$ of arcs such that $v_1 = i$, $v_k = j$ and $\forall l\in [k-1], d(v_l)=v_{l+1}$. By abuse of notation, we may also use this notation to denote the set $\{v_i,i\in[k]\}$ of voters in the chain of delegations from $i$ to $j$. 
Moreover, we denote by $T_d(i)$ the directed subtree rooted in $i$ in delegation graph $H_d$, where $i$ has out-degree 0 and all other vertices have out-degree 1. 

Delegations are transitive, meaning that if voter $i$ delegates to voter $j$, and voter $j$ delegates to voter $k$, then voter $i$ indirectly delegates to voter $k$. If an agent votes, she is called the \textit{guru} of the people she represents and has an \emph{accumulated voting weight} equal to the total weight of people who directly or indirectly delegated to her. 
We denote by $d^{*}_i$ the guru of voter $i$ and by $Gu_d=\{i \in V | d(i)=i\}$ the set of gurus induced by $d$, i.e., the roots of the in-trees in $H_d$. 
%The voter who votes on behalf of voter $i$ is called the \textit{guru} of $i$ and denoted by in $H_d$. 
If an agent delegates, she is called a \textit{follower} and has an accumulated voting weight worth  $0$. Hence, the delegation function $d$ induces an \emph{accumulated weight function} $\alpha_d$, such that $\alpha_d(i) = \sum_{j \in T_d(i)} w_j$ if $i\in Gu_d$ and 0 otherwise.

To define DSGs, Zhang and Grossi defined another weight function.   
Given a set $C \subseteq V$, let us denote by $T_d(i,C)$ the directed subtree rooted in $i$ in the delegation graph $H_d[C]$. 
The subtree $T_d(i,C)$ contains each voter $h$ such that $d^*_h = i$ and $c_d(h,i)$ only contains elements from $C$. 
Given a set $C \subseteq V$, we define the weight function $\gamma_{d,C}$, such that $\gamma_{d,C}(i) = \sum_{j \in T_d(i,C)} w_j$ if $i\in Gu_d$ and 0 otherwise.
Moreover, we denote by $\gamma_d(C) = \sum_{i\in C} \gamma_{d,C}(i)$. The value $\gamma_d(C)$ represents the sum of weights of voters that have a guru in $C$ and such that the chain of delegations leading to this guru is contained in $C$.  
%For $C \subseteq V$, we denote by $d^*_C(h) = j$ the fact that $j$ is the guru of $h$ and that $c_d(h,j)$ only contains elements from $C$. 
%We write $\hat{D}(C) = \{j\in C| \exists k\in C, d^*_C(j) = k\}$ the set of voters that have a guru in $C$ and such that the chain of delegations leading to this guru is contained in $C$. 

%\end{itemize}
%Agents can vote themselves (in this case they are called gurus) or delegate to another agent (in this case they are called delegators). $d_i = j\neq i$ if $i$ delegates to $j$ and $d_i = i$ if voter $i$ votes.
%We write $d=(d_1, d_2, ..., d_n)$ to denote a \textit{delegation profile}, representing each voter's delegation choice, i.e., $d_i=d(i)$ for any $i \in [n]$.

\subsection{Delegative Simple Games}
We call a Liquid Democracy Election (LDE) a tuple $\mathcal{E} = \langle D, \omega, d, q \rangle$ or $\mathcal{E} = \langle D, \omega, H_d, q \rangle$ (we may use $H_d$ in place of $d$ when convenient). 
DSGs are motivated by LDEs. In the DSG $\mathcal{G}_{\mathcal{E}}$ induced by an LDE $\mathcal{E} = \langle D, \omega, d, q \rangle$, $\nu_{\mathcal{E}}(C) = 1$ iff $\gamma_{d}(C) \ge q$, i.e., a coalition $C$ is winning whenever the sum of weights accumulated by gurus in $C$ from agents in $C$ meets the quota. %The delegative Banzhaf index of an agent $i$ is defined as the Banzhaf index of $i$ in the delegative simple game of $V$, $DB_i(V) = B_i(G_V)$.  
This defines a new class of simple games which have a compact structure but different from the one of WVGs. 
In our case, the weights are not given in advance, but rather derived from a graph structure. 
The different power indices defined for simple games can of course be applied to this new class.  
It is worth noticing that in DSGs the power of each voter $i$ does not only depend on the amount of delegations she receives through delegations, but also on the structure of the subtree rooted in her in the delegation graph. Notably, the more direct the supports of $i$ are, the more important she is in the delegation graph (see Property~\ref{prop : axiom MI} for a formal statement).

\begin{definition}[Active Agent]\label{defActiveAgent}
Consider a delegation function $d$ and a coalition $C \subseteq V$. We say that voter $i \in C$ is \emph{active} in $C$ if $d^*(i) \in C$ %\in \hat{D}(C)$ $j$ is the guru of $h$ 
and $c_d(i,d^*(i)) \subseteq C$; 
otherwise, $i$ is called an \emph{inactive} agent. 
\end{definition}

Notice that according to the definition of DSGs, an active (resp. inactive) agent $i \in C$ contributes weight $w_i$ (resp. $0$) to the coalition $C$ (see Example~\ref{exNotations}). Given a DSG $\mathcal{G}_{\mathcal{E}}$, let $DB_i(\mathcal{E})$ and $DS_i(\mathcal{E})$ denote the delegative Banzhaf and Shapley-Shubik values of voter $i$ in $\mathcal{G}_{\mathcal{E}}$, respectively. 
%Sometimes we may write $B_i(\mathcal{G}_{\mathcal{E}})$ (resp. $S_i(\mathcal{G}_{\mathcal{E}})$) instead of $DB_i(\mathcal{E})$ (resp. $DS_i(\mathcal{E})$). 
When speaking about these indices, we drop parameter $\mathcal{E}$ (or $\mathcal{G}_{\mathcal{E}}$) when it is clear from the context. %These values are referred to as the delegative Banzhaf and Shapley-Shubik indices of voter $i$. 
Let us give an illustrative example to understand all aspects of our model.

\begin{exmp}\label{exNotations}
Consider an LDE $\mathcal{E} = \langle D = (V , A), \omega, d, q \rangle$, where $V=[8]$ is a set of $8$ voters delegating through a SN with $d(1) = d(2) = d(3) = 3$, $d(4) = d(6) = 7$, $d(5) = 6$ and $d(7) = d(8) = 8$ as illustrated in Fig.~\ref{figNotation} and $q=3$. Each voter $i$ has weight $\omega(i) = 1$. The set of gurus is $Gu_d=\{3, 8\}$, so, $\alpha_d(3)=\sum_{j \in T_d(3)}w_j=3$, $\alpha_d(8)=\sum_{j \in T_d(8)}w_j=5$, and for any $i \in V\setminus\{3,8\}$, $\alpha_d(i)=0$. Consider the set $C=\{3,5,7,8\}$. $T_d(8,C)$ (resp. $T_d(3,C)$) is the subtree rooted  in $8$ (resp. $3$) and  composed of voters $\{7,8\}$ (resp. $\{3\}$). Thus, $\gamma_{d,C}(3)=\sum_{j \in T_d(3, C)}w_j=1$, $\gamma_{d,C}(8)=\sum_{j \in T_d(8, C)}w_j=2$ and $\gamma_{d,C}(i)=0$ for $i \in \{5,7\}$, and then $\gamma_d(C)=\sum_{i \in C} \gamma_{d,C}(i)=3$. 
Also, %$d^*_C(3)=3$, $d^*_C(7)=8$, $d^*_C(8)=8$ and $\hat{D}(C)=\{3,7,8\}$ as for any $i \in \{3,7,8\}$, $c_d(i,d^*_C(i)) \subseteq C$
the set of active agents is $\{3, 7, 8\}$. Note that $5$ % \notin \hat{D}(C)$ 
is an inactive agent in $C$ as $c_d(5,8) \not \subseteq C$. Now we intend to compute $DB_8$ and $DB_6$. 

We first compute $DB_8$ by counting the coalitions $C \subseteq V\setminus\{8\}$ for which $8$ is a swing agent. 
Note that if $\{1,2,3\}\subseteq C$, then $8$ cannot be a swing agent for $C$. 
Second, if $\{1,2,3\}\cap C\in\{\{1,3\},\{2,3\}\}$, then $8$ will always be a swing agent for $C$. There are 32 such coalitions. 
Third, if $\{1,2,3\}\cap C = \{3\}$, then $8$ will be a swing agent for $C$ iff $7\in C$. There are 8 such coalitions. 
Last, if $3 \not\in (\{1,2,3\}\cap C)$, then $8$ will be a swing agent for $C$ iff $|C\cap\{4,6,7\}| \ge 2$. There are 24 such coalitions.
Hence $DB_8(V) = \frac{1}{2^{7}}\times 64=\frac{1}{2}$.

Now we compute $DB_6$. Note that in order for $6$ to be a swing agent in any coalition $C$, it is necessary that $\{7,8\}\subseteq C$. 
Hence, the coalitions $C$ for which $6$ is a swing agent are of the form $C = \{7,8\}\cup S$ where $S\subseteq \{1,2,5\}$. The number of such coalitions is 8. Hence, $DB_6=\frac{1}{2^{7}}\times 8=\frac{1}{16}$.

\begin{figure}[!h]
		\centering
		\begin{tikzpicture}
				\tikzset{vertex/.style = {shape=circle,draw = black,thick,fill = white}}
				\tikzset{edge/.style = {->,> = latex'}}
				\tikzset{edgebis/.style = {--}}
				
				\node[vertex] (8) at  (1,2.5) {$8$};
				\node[vertex] (7) at  (0,2) {$7$};
				\node[vertex] (6) at  (-1,1.6) {$6$};
				\node[vertex] (5) at  (-2,1) {$5$};
				\node[vertex] (4) at (.5,1) {$4$};
				
				\node[vertex] (3) at  (3,2.5) {$3$};
				\node[vertex] (2) at  (2.5,1.5) {$2$};
				\node[vertex] (1) at  (3.5,1.5) {$1$};

				%edges
				\draw[edge] (7) to (8);
				\draw[edge] (6) to (7);
				\draw[edge] (5) to (6);
				\draw[edge] (4) to (7);
				
				\draw[edge] (1) to (3);
				\draw[edge] (2) to (3);
				
				\draw[edge,dotted,bend right] (6) to (5);
				\draw[edge,dotted,bend right] (7) to (4);
				\draw[edge,dotted] (2) to (7);
				\draw[edge,dotted,bend right] (3) to (2);
				\draw[edge,dotted] (3) to (8);
				
				\draw[edge,dotted,bend left] (1) to (4);
		\end{tikzpicture}

		\caption{ The delegation graph $H_d$ (dotted arcs are the other arcs of $D = (V,A)$) with the set of gurus $Gu_d=\{3,8\}$.} \label{figNotation}
	\end{figure}
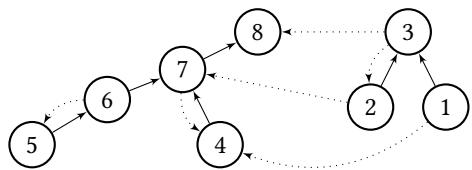
\end{exmp}

%In the next section, we investigate several %axiomatic and computational 
%properties of the delegative Banzhaf measure and the delegative Shapley-Shubik index. %address the computational problem of computing the delegative Banzhaf and Shapley-Shubik indices of a given voter $i$. Note that all missing proofs are deferred to the Appendices.

\section{Properties of DB and DS} \label{section : properties}

This section investigates properties of the delegative Banzhaf measure and the delegative Shapley-Shubik index. First we investigate if, these values can be computed efficiently. 

%\ED{Most part of the next paragraph are repeated in the Related Work section. Maybe we could briefly mention that the standard version is hard and refer the reader to the Related Work section.} 

Evaluating the ``standard'' Banzhaf and Shapley-Shubik measures in WVGs is \#P-complete~\cite{deng1994complexity,prasad1990np} and several decision problems related to their computation are NP-complete or coNP-complete~\cite{chalkiadakis2011computational,matsui2001np,prasad1990np}. 
As DSGs include WVGs as a subcase (consider the restriction when each voter votes herself), these hardness results also hold for DSGs. 
However, these measures can be computed by a pseudo-polynomial algorithm~\cite{matsui2001np} in WVGs, that is an algorithm running in time $poly(n, w_{\max})$, where $w_{\max}:=\max_{i\in V} \omega(i)$ is the maximal weight of an agent. 
We can extend this dynamic programming approach to compute the delegative Banzhaf measure and the delegative Shapley-Shubik index in pseudo-polynomial time.

%\parbox{11cm}{
\begin{restatable}{theorem}{thShBaSudoPolynomial}\label{thShBaPolynomial}
Given a delegative simple game $\mathcal{G}_{\mathcal{E}}$ induced by an LDE $\mathcal{E} = \langle D = (V,A), \omega, d, q \rangle$ and a voter $i \in V$, $DS_i$ and $DB_i$ can be computed in $O(n^4w_{\max}^2)$ and $O(n^2 w_{\max}^2)$, respectively. 
\end{restatable}

The description of the algorithm given in the Appendix %Appendix~\ref{App : properties}
is similar to the one used in WVGs but it carefully takes into account the delegation graph provided in the input. All missing or incomplete proofs can be found in the Appendix.%Appendices~\ref{App : properties}, \ref{app : hardness} and \ref{app : hardness2}.

We now investigate how these measures evolve when the delegation structure of the LDE is slightly modified. The purpose for studying these properties is that they are notably relevant when studying bribery problems where one tries to modify the structure of the instance to increase (or decrease) the relative importance of a given agent, as will be done in Section~\ref{section : bribery}. Possible modifications of the delegation structure can consist in adding a direct or indirect delegation from one agent to another.% or splitting an agent in two. For this latter type of modification, we introduce the notion of bloc formation from Zhang and Grossi~\cite{zhang2020power}.  

We now present three properties. Let $\nu_{\mathcal{E}}$ be the characteristic function of a DSG $\mathcal{G}_{\mathcal{E}}$ induced by an LDE $\mathcal{E} = \langle D = (V , A), \omega, d, q \rangle$. In the following $f$ denotes a generic delegative power measure and $f_i$ is the delegative power measure value of agent $i$. %We may write $\mathcal{E} = \langle D = (V , A), \omega, d, q \rangle$ instead of $\mathcal{E} = \langle D = (V , A), \omega, d, q \rangle$ if there is no confusion.%involved in our characterization.

%\begin{axiom}[Dummy Player (\textbf{DP})]\label{axiomDP}
%If $i$ is a dummy player, then $f_i(\mathcal{E})=0$.
%\end{axiom}

%\begin{axiom}[Efficient Index (\textbf{EI})]\label{axiomEI}
%For any LDE $\mathcal{E}$, $\sum_{i\in V}f_i(\mathcal{E})=\nu_\mathcal{E}(V)$.
%\end{axiom}

%\begin{axiom}[Symmetric Agents (\textbf{SA})]\label{axiomSA}
%If agents $i$ and $j$ are symmetric, $f_i(\mathcal{E})=f_j(\mathcal{E})$.
%\end{axiom}

The following property states that if some voters delegate to a voter $i$ directly, $i$ should get more powerful.
\begin{axiom}[Monotonicity w.r.t. Direct Delegations (\textbf{MDD})]
Consider two LDEs $\mathcal{E} = \langle D = (V , A), \omega, d, q \rangle$ and $\mathcal{E}' = \langle D = (V , A), \omega, d', q \rangle$ and two agents $i,j \in V$ such that $j \not\in  c_d(i,d_i^*)$, $d'(j)=i$ and $d(l) = d'(l)$ for $l\neq j$. Then, $f_i(\mathcal{E}') \ge f_i(\mathcal{E})$.
\end{axiom}

Next, we introduce another property which says that if a voter receives some delegation (directly or indirectly), she should become more powerful.
\begin{axiom}[Monotonicity w.r.t. Delegations (\textbf{MD})]
Consider two LDEs $\mathcal{E} = \langle D = (V , A), \omega, d, q \rangle$ and $\mathcal{E}' = \langle D = (V , A), \omega, d', q \rangle$ and three agents $i,j,k \in V$ s.t. $j \not\in  c_d(k,d_k^*)$, $i \not\in  c_d(j,d_j^*)$, $i \in  c_d(k,d_k^*)$, $d'(j)=k$ and $d(l) = d'(l)$ for $l\neq j$. Then, $f_i(\mathcal{E}') \ge f_i(\mathcal{E})$.
\end{axiom}

Consider a delegation graph $H_d$ and two voters $i, j \in V$ such that $i \in c_d(j, d^*_{j})$. We now introduce the last property stating that in a new delegation graph $d'$ obtained from $d$ in which all voters have the same delegation strategies as $d$ except that $j$ delegates to a voter $l \in c_d(j, d^*_j) \setminus c_d(i, d^*_i)$, $i$ gets more powerful. In other words, this property says that if the voters who support a specific voter $i$ through some intermediaries, support $i$ through a smaller set of intermediaries, $i$ gets more powerful.

\begin{axiom}[Monotonicity w.r.t. Intermediaries (\textbf{MI})]\label{prop : axiom MI}
Consider two LDEs $\mathcal{E} = \langle D = (V , A), \omega, d, q \rangle$ and $\mathcal{E}' = \langle D = (V , A), \omega, d', q \rangle$ and three agents $i,j, k \in V$ such that $k \in c_d(j, d^*_j)$, $i \in c_d(k, d^*_k)$, $\forall l\in V\setminus \{j\}, d(l)=d'(l)$ and $d'(j)=k$. Then, $f_i(\mathcal{E}') \ge f_i(\mathcal{E})$. 
\end{axiom}

%Before proceeding, we note that there are different ways in which the Shapley value of an agent can be expressed. We recall the following one. Let $\Pi[n]$ (or $\Pi$ when the value $n$ is clear from the context) be all possible permutations of voters in $[n]$. For any $\pi \in \Pi$, let $C_{\pi}(i)$ denote the set of all voters preceding i in $\pi$. Let $m^{\nu_\mathcal{E}}_{\pi}(i)$ be the marginal contribution of voter $i$ with respect to $\pi \in \Pi$ which is defined as follows:
%\begin{align}\label{alignMarginalContribution}
 %   m^{\nu_\mathcal{E}}_{\pi}(i) := \nu_\mathcal{E}(C_{\pi}(i)\cup \{i\})-\nu_\mathcal{E}(C_{\pi}(i)).
%\end{align}
%When $\nu_\mathcal{E}$ is clear from the context, we write $m_{\pi}(i)$ instead of $m^{\nu_\mathcal{E}}_{\pi}(i)$. It is well known that the Shapley value of voter $i$ can be rewritten as follows:
%\begin{align}\label{alignShapelySecond}
 %   DS_i=\frac{1}{n!}\sum_{\pi \in \Pi} m_{\pi}(i).
%\end{align}
%Recall that in a delegative simple game $\mathcal{G}_{\mathcal{E}}$, $\nu_\mathcal{E}(C)=1$ iff $\sum_{i\in \hat{D}(C)}\omega(i)\ge q$, where $\hat{D}(C)$ represents the set of all active voters in $C$.

We now investigate which of these three properties are satisfied by the delegative Banzhaf and Shapley-Shubik measures. Counter-intuitively, we first show that while these indices satisfy properties \textbf{MDD} and \textbf{MI}, it is not the case for property \textbf{MD}.
\begin{proposition}\label{prMonotonicityDirectDelegation}
The delegative Banzhaf measure and the delegative Shapley-Shubik index satisfy %\textbf{DP}, \textbf{EI}, \textbf{SA}, 
property \textbf{MDD}.
\end{proposition}
\begin{proof}
%We now turn to property \textbf{MDD}. 
Consider two LDEs $\mathcal{E} = \langle D = (V , A), \omega, d, q \rangle$ and $\mathcal{E}' = \langle D', \omega', d', q' \rangle$ and two agents $i,j \in V$ such that $j \not\in  c_d(i,d_i^*)$,  $D=D'$, $\omega = \omega'$, $q=q'$, $d'(j)=i$ and $d(l) = d'(l)$ for $l\neq j$. Let us consider a coalition $C\subseteq V\setminus\{i\}$ such that $\nu_{\mathcal{E}}(C \cup\{ i \}) - \nu_{\mathcal{E}}(C) = 1$. This implies that $c_d(i,d_i^*)\subseteq C\cup\{i\}$. 
From $\nu_{\mathcal{E}}(C \cup\{ i \}) = 1$, it is straightforward that $\nu_{\mathcal{E}'}(C \cup\{ i \}) = 1$ as $d(l) = d'(l)$ for $l\neq j$, $d'(j)=i$ and $c_d(i,d_i^*)\subseteq C\cup\{ i \}$. Moreover, from $\nu_{\mathcal{E}}(C) = 0$, it is also easy to see that $\nu_{\mathcal{E}'}(C) = 0$ as $d(l) = d'(l)$ for $l\neq j$, $d'(j)=i$ and $i\not\in C$. 
Because this reasoning holds for any coalition for which $i$ is a swing agent, we obtain that  $DS_i(\mathcal{E}') \ge DS_i(\mathcal{E})$ and $DB_i(\mathcal{E}') \ge DB_i(\mathcal{E})$. 
\end{proof}

\begin{proposition}\label{obs:md}
The delegative Banzhaf measure and the delegative Shapley-Shubik index do not satisfy property \textbf{MD}.
\end{proposition}
\begin{proof}
Consider an LDE $\mathcal{E}=\langle D=(V,A), \omega, d,q\rangle$, where $V=\{1,2,3\}$ is a set of $3$ agents delegating through a complete SN $D$ ($A$ contain all possible arcs) with delegations $d(2)=d(1)=1$ and $d(3)=3$, all weights equal to one, and $q=2$. In $\mathcal{E}$ voter $1$ is a swing agent for sets in $\{\{2\},\{3\},\{2,3\}\}$. Now consider the LDE $\mathcal{E}'$ which is identical to $\mathcal{E}$ except that $d(3)=2$. In $\mathcal{E}'$ voter $1$ is a swing agent for sets in $\{\{2\},\{2,3\}\}$. Hence, $DS_1(\mathcal{E}') < DS_1(\mathcal{E})$ and $DB_1(\mathcal{E}') < DB_1(\mathcal{E})$.
\end{proof}
Hence, Proposition~\ref{obs:md} highlights a paradox for measures $DB$ and $DS$: a voter who receives more voting power can become less powerful because of the delegation structure underlying the DSG.
Hence, if one studies a bribery problem in which she wants to change some delegations to make an agent $i^*$ more powerful, then she is on the safe side if she decides to add direct delegations to $i^*$ but adding indirect delegations may be counterproductive. 
 
\begin{proposition}\label{prMonotonicityIntermediary}
The delegative Banzhaf measure and the delegative Shapley-Shubik index satisfy property \textbf{MI}.
\end{proposition}
\begin{proof}
One may use the same argument as that of Proposition~\ref{prMonotonicityDirectDelegation} to prove the proposition.% Consider two LDEs $\mathcal{E} = \langle D = (V , A), \omega, d, q \rangle$ and $\mathcal{E}' = \langle D', \omega', d', q' \rangle$ and three agents $i,j, k \in V$ such that $k \in c_d(j, d^*_j)$, $i \in c_d(k, d^*_k)$, $\forall l\in V\setminus \{j\}, d_l=d'_l$ and $d'_j=k$, $D=D'$, $\omega = \omega'$ and $q=q'$. Let us consider a coalition $C\subseteq V\setminus\{i\}$ such that $\nu_{\mathcal{E}}(C \cup\{ i \}) - \nu_{\mathcal{E}}(C) = 1$. This implies that $c_d(i,d_i^*)\subseteq C\cup\{i\}$. 
%From $\nu_{\mathcal{E}}(C \cup\{ i \}) = 1$, it is straightforward that $\nu_{\mathcal{E}'}(C \cup\{ i \}) = 1$ as $d(l) = d'(l)$ for $l\neq j$, $d'(j)=k \in c_d(i,d_i^*)$ and $c_d(i,d_i^*)\subseteq C\cup\{ i \}$. Moreover, from $\nu_{\mathcal{E}}(C) = 0$, it is also easy to see that $\nu_{\mathcal{E}'}(C) = 0$ as $d(l) = d'(l)$ for $l\neq j$, $d'(j)=k \in c_d(i,d_i^*)$ and $i\not\in C$. 
%Because this reasoning holds for any coalition for which $i$ is a swing agent, we obtain that  $DS_i(\mathcal{E}') \ge DS_i(\mathcal{E})$ and $DB_i(\mathcal{E}') \ge DB_i(\mathcal{E})$. 
\end{proof}

\section{Bribery by Delegation Modifications} \label{section : bribery}
\subsection{Power index modification by bribery} \label{section : PIM}
The support structure in DSGs induces the following natural question:  which voters should one influence under a budget constraint to maximize/minimize the voting power of a given voter? 
This question leads to the following computational bribery problems.

\begin{center}
\noindent\fbox{\parbox{14cm}{
		\emph{Problems}: \textbf{BMinP}, \textbf{SMinP}, \textbf{BMaxP} and \textbf{SMaxP}\\
		\emph{Input}: An LDE $\mathcal{E} = \langle D= (V, A), \omega, d , q \rangle$, a voter $v^* \in V$, a budget $k \in \mathbb{N}$, and a threshold $\tau\in\mathbb{Q}_+$.\\
		\emph{Feasible solution}: A delegation function $d'\in \Delta(D)$ s.t. $|\{i\in V : d(i)\neq d'(i)\}|\le k$ leading to an LDE $\mathcal{E}' = \langle D, \omega, d', q \rangle$. \\
		\emph{Question}: Can we find a feasible solution $d'$ such that:
		\begin{center}
		\begin{tabular}{cc}
		     \textbf{BMinP}: $DB_{v^*}(\mathcal{E}') \le \tau$? & \textbf{BMaxP}: $DB_{v^*}(\mathcal{E}') \ge \tau$?\\
		     \textbf{SMinP}: $DS_{v^*}(\mathcal{E}') \le \tau$? & \textbf{SMaxP}: $DS_{v^*}(\mathcal{E}') \ge \tau$?
		\end{tabular}
		\end{center}
}}\\
\end{center}

Stated otherwise, in the Banzhaf Minimization (resp. Maximization) Problem, \textbf{BMinP} (resp. \textbf{BMaxP}) for short, we wish to determine if we can make the Banzhaf measure of voter $v^*$ lower (resp. greater) than or equal to a given threshold, by only modifying $k$ delegations. This cardinality constraint can be justified by the fact that influencing each voter is costly. 
\textbf{SMinP} and \textbf{SMaxP} are similar problems corresponding to the Shapley-Shubik index.  
While \textbf{BMaxP} and \textbf{SMaxP} correspond to the constructive variant of the bribery problem, \textbf{BMinP} and \textbf{SMinP} correspond to its destructive variant. These bribery problems are natural in the setting of LD, where one voter could for instance try to get the delegations of several other voters to increase her influence on the election. Moreover, we believe these bribery problems are also relevant in more traditional elections where several politicians or political parties could seek which alliances to foster as to increase their centrality or to make an opponent powerless.

%\subsection{Problem IMinP}
We first show several hardness and hardness of approximation results on these four problems. %\footnote{All incomplete proofs can be found in the Supplementary Material file.%Appendices~\ref{App : properties}, \ref{app : hardness} and \ref{app : hardness2}.
%}  %are inapproximable on trees. %NP-hard. %Note that computing the delegative Banzhaf or Shapley-Shubik indices in these problems is already intractable unless the voters' weights are given in unary. However, our hardness results hold even if voters' weights are given in unary (resp. even if all voters have weight one) for \textbf{BMaxP} and \textbf{SMaxP} (resp. for \textbf{BMinP} and \textbf{SMinP}).
%\begin{comment}

\begin{restatable}{theorem}{IMinPIsHard}\label{thrm : IMinP is NP-Hard}
    The restriction of \textbf{BMinP} and \textbf{SMinP} to the case where all voters have weight 1, i.e., when $\forall i\in V, \omega(i) = 1$, is NP-complete. Moreover, under the same restriction, the minimization versions of problems \textbf{BMinP} and \textbf{SMinP} cannot be approximated within any factor in polynomial time if $\mathtt{P} \neq \mathtt{NP}$.
\end{restatable}
\begin{proof}[Sketch of proof%((Complete proof in Appendix~\ref{app : hardness})
]
%The complete proof can be found in the Supplementary Material file.
We use a reduction from the NP-complete Hamiltonian path problem~\cite{garey1979computers} where the goal is to determine if there exists a path in an undirected graph that visits each vertex exactly once. From an instance of the Hamiltonian path problem with $n$ vertices we create an instance of \textbf{BMinP} (or \textbf{SMinP}). 
The quota of this \textbf{BMinP} (or \textbf{SMinP}) instance is set such that a specific voter will have power index value 0 iff it is at the end of a delegation path of length $n$. 
The idea is that such a path necessarily consists in a Hamiltonian path in the original instance. 
%where one can make a specific agent powerless (power index of 0) iff there is an Hamiltonian path in the instance we reduce from. 
Hence, in the \textbf{BMinP} (or \textbf{SMinP}) instance a successful solution for the briber will consist in creating a delegation path along a Hamiltonian path (if one exists).  The hardness of approximation result is then obtained from the fact that a polynomial-time algorithm with some multiplicative approximation guarantee would be able to distinguish between instances where we can make the agent dummy and the ones where we cannot.
\end{proof}
%\end{comment}
%We can even show a more stronger result.

\begin{restatable}{theorem}{IMinPIsCoHardOnTrees}\label{thIMinPISCoNPHardOnTrees}
Problems \textbf{BMinP} and \textbf{SMinP} are coNP-hard even if the SN is a tree. %cannot be approximated within any factor in polynomial time if $\mathtt{P} \neq \mathtt{coNP}$ even if the SN is a tree. 
\end{restatable}

We move to problems \textbf{BMaxP} and \textbf{SMaxP}.

\begin{restatable}{theorem}{IMaxPIsHard}\label{thIMaxP}
Problems \textbf{BMaxP} and \textbf{SMaxP} with voters' weights and the quota given in unary are NP-complete.   
\end{restatable}
%\end{comment}

%Next we show that \textbf{BMaxP} and \textbf{SMaxP} are inapproximable on trees too.

\begin{theorem}\label{thInApproximableMax}
The maximization versions of problems \textbf{BMaxP} and \textbf{SMaxP} cannot be approximated within any factor in polynomial time if $\mathtt{P} \neq \mathtt{NP}$ even if the SN is a tree. 
\end{theorem}
\begin{proof}
Consider the subset sum problem with positive integers. 
An instance $J$ of this problem is composed of a set of positive integers $S=\{a_1, \dots, a_n\}$ and a target sum $M$: $J=\langle S, M \rangle$ is a Yes-instance iff there exists a subset $L \subseteq S$ such that $\sum_{a_i \in L} a_i=M$. 
We transform an instance $J=\langle S, M\rangle$ of the subset sum problem to an instance $I=\langle \mathcal{E} = \langle D= (V, A), \omega, d , q \rangle , v^*, k, \tau \rangle$ of \textbf{BMaxP} (resp. \textbf{SMaxP}). 
We create a tree $D$ where: 
\begin{itemize}
    \item $V = V_1 \cup \{v,v',v^*\}$ with $V_1 = \{u_i : a_i \in S\}$. 
    \item $ A = \{(v_i, v): v_i \in V\setminus\{v\}\}$. 
\end{itemize}
Let $R = \sum_{a_i \in S} a_i$ and $q = R +2M +3$. The weight function $\omega$ is set as follows: $\omega(v')=R+M+2$, $\omega(v^*)=1$, $\omega(v)= M + 1$ and for any $u_i \in V_1$, $\omega(u_i) = a_i$. 
The initial delegation function $d$ is set as follows: $d(v')=v'$, $d(v^*)=v^*$, $d(v)=v$ and for any $u_i \in V_1$, $d(u_i)=v$. % and $d_{v_j}=v_j$ for any $v_j \in V_2$. 
As $v^*$ is not swing in any coalition $C$ for $\mathcal{E}$, $DB_{v^*}(\mathcal{E})=0$ (resp. $DS_{v^*}(\mathcal{E})=0$). 
First notice that all successful coalitions should contain voter $v'$ and that $v^*$ cannot be a swing agent in a coalition that contains both $v$ and $v'$. In fact, to make $v^*$ a swing for some coalition, one should select a subset $U$ of voters from $V_1$ such that $\sum_{v \in U} \omega(v) = M$ and remove their delegations from $v$. Indeed, $v^*$ is then a swing agent for the coalition $U\cup\{v'\}$.

Now, assume, for a contradiction, that there is a (not-necessarily constant) factor $\beta$ ($0 <\beta \le 1$) polynomial time approximation algorithm, $\mathcal{A}$, for \textbf{BMaxP} and \textbf{SMaxP}. 
Imagine using $\mathcal{A}$ several times with $k = 1$ to $k = n$, where $n=|S|$ in the subset sum instance. 
Thus, if $J$ is a Yes-instance, then there exists a set $L \subseteq S$ such that $\sum_{a_i \in L} a_i=M$. Consider, one of minimal size $k_{min}$. Then, for $k = k_{min}$, $\mathcal{A}$ will necessarily output a delegation function to \textbf{BMaxP} (resp. \textbf{SMaxP}) for which $DB_{v^*}(\mathcal{E}') > 0$ (resp. $DS_{v^*}(\mathcal{E}') > 0$), and in which the set of delegations changed affects the voters $\{u_i:i\in L\}$. By investigating the solution, one can check in polynomial time if it indeed corresponds to a valid certificate for the subset sum problem.
%Conversely, if $J$ is a No-instance, $\mathcal{A}$ will never return a delegation must return a solution to \textbf{BMaxP} (resp. \textbf{SMaxP}) in which $DB_{v^*}(\mathcal{E}')=0$ (resp. $DS_{v^*}(\mathcal{E}')=0$). We see that based on these two cases, the algorithm $\mathcal{A}$ can be used to decide whether $I'$ is a ``yes''-instance or not, which is impossible, unless $P=NP$. 
This concludes the proof.
\end{proof}

More positively, when the input is a complete graph, we can provide more positive results on the approximation viewpoint. We detail an algorithm, called GAMW, standing for Greedy Algorithm with Maximum Weight, which works as follows.

\textbf{If $v^*$ is a guru.}
It iteratively picks a guru $g$ (different from $v^*$) with maximum accumulated weight (if there are several gurus with the same accumulated weight, it selects one arbitrarily), set $d(g)=v^*$ and $k=k-1$. This process continues until $k=0$ or no guru remains to delegate to $v^*$.

\textbf{If $v^*$ is a follower.} Let $Del_d(j) = c_d(j,d^*(j))\setminus\{j\}$ be the set of voters to which $j$ delegates to directly or indirectly. GAMW distinguishes two subcases: i) $k \ge 2$, it assumes that $v^*$ is a guru and sets $d(v^*) = v^*$ and $k=k-1$. It then proceeds as when $v^*$ is a guru, ii) $k=1$, GAMW checks if $q-\sum_{i \in Del_{d_1}(v^*)}\omega(i) > 0$ (i.e., otherwise she is a dummy player.), then it finds a voter among the gurus $g\neq d_{v^*}^*$ and the voters who delegate directly to $Del_d(j)$ with the highest accumulated weight and make her delegate to $v^*$, otherwise (i.e., $q-\sum_{i \in Del_{d_1}(v^*)}\omega(i)\le 0$) it sets $d(v^*) = v^*$.
%i) Similarly as before, in the first case, we find a guru $g\neq d_{v^*}^*$ with the highest weight and make her delegate to $v^*$. Let $H_{d_1}$ be the resulting delegation graph and $W_1=\sum_{i \in T_{d_1}(v^*)} \omega(i)$ be $v^*$'s weight after this process. ii) The second case assumes that $v^*$ is a guru and sets $d(v^*) = v^*$ and $k=k-1$. It then proceeds as when $v^*$ is a guru. Let $H_{d_2}$ be the resulting delegation graph and $W_2=\sum_{i \in T_{d_2}(v^*)} \omega(i)$ be $v^*$'s weight in this case. If $q-\sum_{i \in Del_{d_1}(v^*)}\omega(i) > 0$ and $W_1- \sum_{i \in Del_{d_1}(v^*)}\omega(i)> W_2$ then GAMW considers $H_{d_1}$ as the final solution; otherwise it considers $H_{d_2}$ as the final solution.

\begin{comment}
Let $t_{v^*}(r) \subseteq t_{v^*}$ be a subtree of size $r$ rooted in $v^*$, where $r \in \{1,\dots, |t_{v^*}|\}$. Keep in mind that all voters have the same weight. Let $\subseteq V \setminus t_{v^*}$ be the loosing coalition with the maximum weight. Let $t_{v^*}(r')$ be the subtree with the minimum size such that $C'=C\cup t_{v^*}(r')$ is a winning coalition. As GAMW assigns a subtree with the highest weight to $v^*$ among all algorithms, respecting the budget constraint, if there is no coalition $C'$ for which $v^*$ is swing in after applying GAMW, no algorithm can result in a better solution. Then we assume that $r' \in \{1,\dots, |t_{v^*}|\}$. Let $H_{d'}$ be the resulting delegation graph after applying GAMW and $s=|T_{d'}(v^*)|$.% be the number of voters who follow $v^*$ (including $v^*$ herself) after applying GAMW.
\end{comment}

\begin{theorem}\label{thApproximationAlgorithmMaximumWeightDifferentWeight}
 GAMW is a factor $\frac{1}{2^{n-1}}$ (resp. $\frac{1}{n!}$) approximation algorithm for \textbf{BMaxP} (resp. \textbf{SMaxP}) on complete graphs.
\end{theorem}

\begin{proof}[Sketch of proof]
The intuition is that, as GAMW assigns a set of subtrees with the highest weight to $v^*$ among all algorithms respecting the budget constraint, if there is no coalition $C'$ for which $v^*$ is swing in after applying GAMW, no algorithm can result in a better solution. In particular, suppose that we are given a delegation graph $H_d$ and a guru $g_{max} \in Gu_d$ with the highest accumulated weight. Consider any losing coalition $C \subseteq V\setminus T_d(g_{max})$ in $H_d$. If the coalition $C'=C \cup T_d(g_{max})$ is not a winning coalition, then no guru $g \in Gu_d$ can make $C$ a winning coalition as $\alpha_d(g_{max}) \ge \alpha_d(g)$. In case $k=1$ and $v^*$ is a follower, we consider any losing coalition $C \setminus T_d(v^*)$, where $Del_d(j) \subseteq C$ and use a similar argument.
\end{proof}

To conclude this subsection, we note that, as shown by Theorems~\ref{thrm : IMinP is NP-Hard},~\ref{thIMinPISCoNPHardOnTrees}, \ref{thIMaxP}, and~\ref{thInApproximableMax}, problems \textbf{BMinP}, \textbf{SMinP}, \textbf{BMaxP} and \textbf{SMaxP} are hard. We also believe that these problems are complex in the sense that the power measures they rely on can be hard to grasp for people not used to solution concepts from cooperative game theory. In the next subsection, instead of maximizing a power  measure, we study a problem with a conceptually simpler objective as surrogate.

\subsection{Voting weight modification by bribery}~\label{section : VWM}
%As shown by Theorems~ \ref{thInApproximableMin} and~\ref{thInApproximableMax}, the optimization versions of problems \textbf{BMinP}, \textbf{SMinP}, \textbf{BMaxP} and \textbf{SMaxP} are inapproximable even on trees. We also believe that these problems are complex in the sense that the power indices they rely on can be hard to grasp for people not used to solution concepts from cooperative game theory. In this subsection, instead of maximizing the power index of a voter, we study a problem with a related but conceptually simpler objective. Indeed
In this subsection, we investigate if we can modify at most $k$ delegation choices to make the accumulated weight of a given voter $i^*$ greater than or equal to a given threshold $\tau$. 
%A voter could indeed be interested in obtaining a given accumulated weight $\tau$. 
%Possible interesting values for $\tau$ can be $q$ or $\sum_{i\in V} \omega(i) - q + 1$. In the first case, $i^*$ and her supporters do not need the support of anyone else to reach the election quota. In the second case, $i^*$ is sure that she has enough weight to veto the election.
We term this optimization problem \textbf{WMaxP} for Weight Maximization Problem. It is clear that problem \textbf{WMaxP} is related to problems \textbf{BMaxP} and \textbf{SMaxP} in the sense that a greater voting weight may result in a greater power measure value. 
However, it is well known from the literature on WVGs that this relation is limited as voters with sensibly different weights may have the same relative importance in the election. 
Less intuitively, it is even possible that if the given voter receives too much weight, we may end up in a situation where the voter's delegative power gets decreased (see Proposition~\ref{obs:md}).

We now formally introduce \textbf{WMaxP}. As this problem does not require to know the quota, we define a Partial LDE (PLDE) as a tuple $\mathcal{E} = \langle D= (V, A), \omega, d \rangle$, i.e., an LDE without a quota value.

\begin{center}
\noindent\fbox{\parbox{14cm}{
		\emph{Problems}: \textbf{WMaxP}\\
		\emph{Input}: A PLDE $\mathcal{E} = \langle D= (V, A), \omega, d \rangle$, a voter $i^* \in V$, a budget $k\in\mathbb{N}$, and a threshold $\tau \in \mathbb{N} $.\\
		\emph{Feasible Solution}: A delegation function $d'\in \Delta(D)$ s.t. $|\{i\in V : d(i)\neq d'(i)\}|\le k$ leading to a PLDE $\mathcal{E}' = \langle D, \omega, d'\rangle$. \\
		%\emph{Feasible Solution}: A delegation graph $H^{'} = (V, A^{'})\in \Delta(D)$ corresponding to a delegation function $d'$ s.t. $|A^{'} \setminus A|\le k$. \\
		\emph{Question}: Can we find a solution $d'$ such that $\alpha_{d'}(i^*) \ge \tau$. 
		%\emph{Question}: Does there exist  delegation graph $H' = (V, A^{'})\in \Delta(G)$ s.t. $|A^{'} \setminus A|\le k$ and $u\in Gu_{d'}$ such that $w_{d'}(u) \ge \tau$.
}}\\
\end{center}

We first show that \textbf{WMaxP} is NP-complete and that \textbf{OWMaxP}, the optimization variant of \textbf{WMaxP}, where the goal is to maximize $i^*$'s weight, cannot be approximated in polynomial time with an approximation ratio better than $1-1/e$ if $P\neq NP$. 
The approximation preserving reduction used is from the maximum coverage problem.% and can be found in the Supplementary Material file. %Appendix~\ref{app : hardness2}.

\begin{restatable}{theorem}{thWTPPreservingFactor}\label{thWTP}
	\textbf{WMaxP} is NP-complete and \textbf{OWMaxP} cannot be approximated with an approximation ratio better than $1-1/e$ if $P\neq NP$, even when all voters have weight one.
\end{restatable}	

To obtain more positive results, we consider both the approximation and the parameterized complexity viewpoints.

%Consider the optimization version of \textbf{WMaxP}, where the goal is to maximize $i^*$'s weight, termed \textbf{OWMaxP}. 
\paragraph{An approximation algorithm point of view.} Interestingly, a more general version of \textbf{OWMaxP}, called \textbf{DTO} (Directed Tree Orienteering), has been investigated by Ghuge and Nagarajan~\cite{ghuge2020quasi}. %Now we aim to show that \textbf{OWMaxP} is closely related to \textbf{DTO}, which is studied by Ghuge and Nagarajan~\cite{ghuge2020quasi}. 
In \textbf{DTO}, we are given a directed graph $D=(V, A)$ with edge costs $c:A \rightarrow \mathbb{Z}^{+}$, a root vertex $r^* \in V$, a budget $B \in \mathbb{Z}^{+}$, and a weight function $p:V \rightarrow \mathbb{Z}^{+}$. For any subgraph $G'$ of a given (directed or undirected) graph $G$, let $V(G')$ and $E(G')$ represent the set of nodes and edges in $G'$. The goal is to find an out-directed arborescence $T^*$ rooted at $r^*$ maximizing $p(V(T^*)) = \sum_{v\in V(T^*)} p(v)$ such that $\sum_{e \in E(T^*)} c(e) \le B$. Ghuge and Nagarajan~\cite{ghuge2020quasi} provided a quasi-polynomial time $O(\frac{\log n'}{\log \log n'})$-approximation algorithm, where $n'$ is the number of vertices in an optimal solution. The authors mentioned that this factor is tight for \textbf{DTO} in quasi-polynomial time. It is worth mentioning that Paul et al.~\cite{paul2020budgeted} proposed a $2$-approximation algorithm for the undirected version of \textbf{DTO}.% has been investigated by . %, termed \textit{budgeted prize-collecting minimum spanning tree}, in which we are given an undirected graph and the task is to find a tree $T^*$ containing a root vertex $r^*$ with the maximum weight $\omega(T^*)$ such that $c(T^*) \le B$. Paul et al.~\cite{paul2020budgeted} proposed a 2-approximation algorithm for this case.

Here we show that any approximation algorithm for a restriction of problem \textbf{DTO} can also be used with \textbf{OWMaxP}, preserving
the approximation factor. In particular, consider the particular instances $I'=\langle D'= (V', A'), r^*, c, p, B \rangle$ of \textbf{DTO}, where for any edge $e \in A'$, $c(e)=\{0, 1\}$. For any node $v \in V'$, there exists at most one incoming edge $e$ of cost $c(e)=0$. More importantly, there is no cycle $C$ in $D'$ with the total cost $\sum_{e \in E(C)}c(e)=0$, i.e., there exists at least one edge $e \in C$ with $c(e)=1$. %for any $e \in E(C)$, $c(e)=0$, i.e., $\sum_{e \in C}c(e)=0$. %there exists at least one edge $e \in C$ with $c(e)=1$. 
%Also, for any $S \subseteq V'$, $f(S)=\sum_{i \in S}w_i$, where $w_i$ is the weight of $i \in V'$. %the cost of each outgoing edge is either $1$ or $0$.
We call \textbf{RDTO} this restriction. %of \textbf{DTO}.

%\paragraph{Connections to \textbf{STO}.} 

\begin{theorem}
Consider a parameter $\beta$, where $0<\beta<1$ (not-necessarily constant). The following statements are equivalent:

\begin{enumerate}
    \item[(i)] There is an $\beta$-approximation algorithm for \textbf{RDTO}.\label{alpha-RDTO}
    
    \item[(ii)] There is an $\beta$-approximation algorithm for \textbf{OWMaxP}.\label{alpha-WmaxP}
\end{enumerate}
\end{theorem}

\begin{proof}
To prove that (i) implies (ii), we proceed as follows. Let $I=\langle D= (V, A), \omega, d,i^*, k \rangle$ be an instance of \textbf{OWMaxP}. We suppose that $d_{i^*}=i^*$; otherwise we define a new delegation graph $d'$ such that $d'(i^*)=i^*$, $d'(i)=d(i)$ for any $i \in V \setminus \{i^*\}$ and $k=k-1$. 
Indeed, $i^*$ should be a guru to have a non-zero accumulated weight. 
Let $I'=\langle D'= (V', A'), r^*, c, p, B \rangle$ be an instance of \textbf{RDTO} obtained from $I$ as follows. 
We set $V'=V$, $r^*=i^*$ and $A'=\{(i,j):(j,i) \in A\}$, i.e., we reverse the edges. For any $e=(i,j) \in A'$ in $I'$, if $d(j)=i$ in $I$, $c(e)=0$; $c(e)=1$ otherwise.
%For any $S \subseteq V'$, let $f(S)=\sum_{i \in S}\omega(i)$.
Lastly, $B=k$ and $p(v)=\omega(v)$ for any $v \in V'$. Consider a solution $T^*$ to \textbf{RDTO} on $I'$. We can simply reverse the edges in $T^*$ and obtain a subtree of $D$ rooted in $i^*=r^*$ with cost $c(T^*) \le B=k$ that induces a delegation graph $d'$. We conclude by noticing that any edge $e=(i, j) \in E(T^*)$ either costs $0$ if $d(j)=i$ or $1$ otherwise and $\alpha_{d'}(i^*) = p(V(T^*))$. This concludes the first direction.

Now we show that (ii) implies (i). %, we give a polynomial-time reduction from \textbf{RDTO} to \textbf{OWMaxP} as follows. 
Let $I'=\langle D'= (V', A'), r^*, c, p, B \rangle$ be an instance of \textbf{RDTO}. 
Let $I=\langle D= (V, A), \omega, d,i^*, k \rangle$ be an instance of \textbf{OWMaxP} obtained from $I'$ as follows. 
We set $V'=V$, $r^*=i^*$ and $A=\{(i,j):(j,i) \in A'\}$, i.e., reversing edges. 
For any voter $i \in V \setminus \{i^*\}$ if there exists an incoming edge $e=(j, i) \in A'$ with cost $c(e)=0$, we set $d(i)=j$; $d(i)=i$ otherwise. 
We set $d(i^*)=i^*$. For any $v \in V$, let $\omega(v)=p(v)$. Lastly, we set $k=B$. 
As there exists no cycle $C \in D'$ with $\sum_{e \in C}c(e)=0$ and for any $i \in V'$ there is at most one incoming edge $e$ with cost $c(e)=0$, the resulting delegation graph $H_d$ is feasible. 
Now consider another delegation graph $H_{d'}$ such that $|\{i\in V : d(i)\neq d'(i)\}|\le k$. 
By reversing the edges in subtree rooted at $i^*$ in $H_{d'}$ we get an arborescence $T^*$ in $D'$ that is rooted at $r^*$ with $p(V(T^*)) = \alpha_{d'}(i^*)$.
\end{proof}

We now present a polynomial-time approximation algorithm for \textbf{OWMaxP} to achieve a trade-off between the violation of budget constraint and the approximation factor. Given an undirected graph $G=(V(G), E(G))$, a distinguished vertex $r \in V(G)$ and a budget $B$, where each vertex $v \in V(G)$ is assigned with a prize $p'(v)$ and a cost $c'(v)$. A graph $G$ is called $B$-proper for the vertex $r$ if the cost of reaching any vertex from $r$ is at most $B$. Consider a subtree $T=(V(T), E(T))$ of $G$, where $V(T) \subseteq V(G)$ and $E(T) \subseteq E(G)$. Let $c'(T)=\sum_{v \in V(T)} c'(v)$ and $p'(T)=\sum_{v \in V(T)} p'(v)$. Let $\gamma=\frac{p'(T)}{c'(T)}$ be the prize-to-cost ratio of $T$. Bateni, Hajiaghayi and Liaghat~\cite{bateni2018improved} proposed a trimming process that leads to the following.

\begin{lemma}[Lemma 3 in \cite{bateni2018improved}]\label{lmBateniTrimmingProcess}
Let $T$ be a subtree rooted at $r$ with the prize-to-cost ratio $\gamma$. Suppose the underlying graph is $B$-proper for $r$ and for $\epsilon \in (0, 1]$ the cost of the tree is at least $\frac{\epsilon B}{2}$. One can find a tree $T^*$ containing $r$ with the prize-to-cost ratio at least $\frac{\epsilon \gamma}{4}$ such that $\epsilon B/2 \le c'(T^*) \le (1+\epsilon)B$.
\end{lemma}

We show that Lemma~\ref{lmBateniTrimmingProcess} can be applied to our case. Given an instance $I=\langle D= (V, A), \omega, d, i^*\in V, k \rangle$ of \textbf{OWMaxP}. We create an edge-cost directed graph $D_d=(V_d,A_d)$ respecting the delegation function $d$ as follows: $V_d=V$, $A_d=A$, each vertex $v \in V_d$ is associated with a weight $\omega(v)$ and each edge $e=(i, j) \in A_d$, is associated with a cost $c(e)=0$ if $d(i)=j$, $c(e)=1$ otherwise. $D_d$ is called the \textit{$d$-edge-cost} graph of $D$. Let $V'$ be all vertices in $V_d$ such that the cost of reaching from any node $v' \in V'$ to $i^*$ is at most $k$. 
We call subgraph $D'=(V', A')$ of $D_d$ $k$-\textit{appropriate} for $i^*$ where $A'=V' \times V' \cap A_d$ (we make this definition to avoid confusions between the undirected and directed cases). Consider a subtree $T$ of $D'$. %, where $V'(T) \subseteq V'$ and $A'(T) \subseteq A'$. 
Let $\omega(T)=\sum_{v \in V'(T)} \omega(v)$ and $c(T)=\sum_{e \in A'(T)} c(e)$. Let $\gamma=\frac{\omega(T)}{c(T)}$ be the weight-to-cost ratio of $T$.

\begin{lemma}\label{lmOurTrimmingProcess}
Given an instance $I=\langle D= (V, A), \omega, d, i^*, k \rangle$ of \textbf{OWMaxP}. Consider the $d$-edge-cost graph $D_d$ which is $k$-appropriate for $i^*$. Let $T$ be a subtree of $D_d$ rooted at $i^*$ with the weight-to-cost ratio $\gamma$. Suppose that for $\epsilon \in (0, 1]$ $c(T) \ge \frac{\epsilon B}{2}$. One can find a tree $T^*$ containing $i^*$ with the weight-to-cost ratio at least $\frac{\epsilon \gamma}{4}$ such that $\epsilon B/2 \le c'(T^*) \le (1+\epsilon)B$.
\end{lemma}

\begin{proof}
Let $V(T)$ and $A(T)$ be the set of vertices and edges of $T$. Now we create another subtree $T'=(V'(T), A'(T))$ as follows: 
\begin{itemize}
    \item $V(T') = V(T) \cup V_1$ with $V_1 = \{v_e : e \in A(T)\}$. 
    \item $ A(T') = \{(i, v_e), (v_e, j):  e=(i,j)\in A(T)\}$. 
\end{itemize}

Each vertex $v \in V(T') \cap V(T)$ (resp. $v_e \in V(T') \cap V_1$) is assigned with a prize $p'(v)=\omega(v)$ (resp. $p'(v_e)=0$) and a cost $c'(v)=0$ (resp. $c'(v_e)=c(e)$). Lemma~\ref{lmBateniTrimmingProcess} can be applied to the subtree $T'=(V(T'), A(T'))$, as the trimming process by Bateni, Hajiaghayi and Liaghat~\cite{bateni2018improved} only removes some subtrees of $T'$ to reach the guarentees mentioned in Lemma~\ref{lmBateniTrimmingProcess}. This completes the proof.
\end{proof}

Now we are ready to propose our approximation algorithm for \textbf{OWMaxP}, called VBAMW. 
Given an instance of \textbf{OWMaxP} $I=\langle D= (V, A), \omega, d, i^*, k \rangle$, VBAMW first creates the $d$-edge-cost graph $D_d=(V_d, A_d)$ which is also maximal inclusion-wise $k$-appropriate graph for $i^*$.  %Given a directed graph $D=(V,A)$ and the delegation graph $H_d$. VBAMW first creates another directed graph $D'=(V',A')$ such that $V'=V$, $A'=A$, each vertex $v' \in V'$ is associated with a weight $w(v)=\omega(v)$ and each edge $e'=(i, j) \in A'$, is associated with a cost $c(e')=0$ if $d(i)=j$, $c(e')=1$ otherwise. Then VBAMW creates a $k$-proper graph $D''=(V'', A'')$ for $i^*$.
Now VBAMW finds a spanning arborescence $T=(V(T), A(T))$ of $D_d$ with minimum cost $c(T)$, using Edmonds' algorithm~\cite{edmonds1967optimum}. If $c(T)\le (1+\epsilon)k$, we are done. Suppose it is not the case. Let $\gamma=\frac{\omega(T)}{c(T)}$ be the weight-to-cost ratio of tree $T$. By Lemma~\ref{lmOurTrimmingProcess}, from tree $T$, VBAMW finds another subtree $T^* \subseteq T$ of the cost at most $(1+\epsilon) k$ and the weight-to-cost ratio $\frac{\epsilon\gamma}{4}$. %Note that in our setting the cost of a tree is on its edges. However, Lemma~\ref{lmBateniTrimmingProcess} can be applied to our case too. 

\begin{theorem}\label{thVBAMW}
    VBAMW is a $\frac{\epsilon^2 k}{8n}$ approximation algorithm with the cost at most $(1+\epsilon)k$ for \textbf{OWMaxP}.
\end{theorem}

\begin{proof}
Let $T$ be the spanning arborescence returned by Edmonds' algorithm~\cite{edmonds1967optimum} with weight-to-cost ratio $\gamma$. It is clear that $\omega(T) \ge OPT$, where $OPT$ is the optimum weight to \textbf{OWMaxP}. By Lemma~\ref{lmOurTrimmingProcess}, VBAMW will find another subtree $T^*$ of cost $\epsilon k/2 \le c(T^*) \le (1+\epsilon)k$ and weight-to-cost ratio:
$$\frac{w(T^*)}{c(T^*)} \ge \frac{\epsilon \gamma}{4} \ge \frac{\epsilon \omega(T)}{4 c(T)} \ge \frac{\epsilon}{4c(T)} OPT \ge \frac{\epsilon}{4n} OPT.$$
As $c(T^*) \ge \epsilon k/2$, we have $\omega(T^*) \ge \frac{\epsilon^2 k}{8n} OPT$, concluding the proof.
\end{proof}

\begin{remark}
VBAMW is a $\frac{\epsilon^2 B}{8B'}$ approximation algorithm with the cost at most $(1+\epsilon)B$ for \textbf{DTO}, where $B'$ is the minimum cost a spanning tree of the given graph can have. To the best of our knowledge, VBAMW is the first polynomial-time approximation algorithm for \textbf{DTO}.
\end{remark}

%\input{SAGT/Contribution/Modification/DTO}
%\HG{End of not checked}

\paragraph{A parameterized complexity point of view.} We now define the two following parameters: 
\begin{itemize}
    \item We denote by $\bar{\mathtt{req}} = \sum_{i\in V} \omega(i) - \tau$ the amount of voting weight that $i^*$ does not need to reach the threshold $\tau$;
    \item We denote by $\mathtt{req} = \tau - \alpha_d(i^*)$ the amount of additional voting weight that $i^*$ needs to reach the threshold $\tau$.
\end{itemize} 
We study the parameterized complexity of \textbf{WMaxP} w.r.t. these two parameters. It can indeed be expected that the problem becomes easier if one of them is small. If $\bar{\mathtt{req}}$ is small, then the combinations of voters that may not delegate to $i^*$ in a solution $d$, such that $\alpha_d(i^*)\ge \tau$, are probably limited.  Conversely, if $\mathtt{req}$ is small, then the number of voters that $i^*$ needs an additional support of to reach $\tau$ is small. These intuitions indeed yield positive results (Theorems~\ref{thrm: XP for req bar} and \ref{thrm: FPT for req}). These two parameters seem to be opposite from one another. Indeed a small value for parameter $\bar{\mathtt{req}}$ (resp. $\mathtt{req}$) indicates that reaching the threshold $\tau$ is probably hard (resp. easy). Parameter $\bar{\mathtt{req}}$ could for instance be small if $\tau = q$ and the election is conservative (i.e., $q$ is close to $\sum_{i \in V} \omega(i)$). Meanwhile, parameter $\mathtt{req}$ can be small if %$\tau = \sum_{i\in V} \omega(i) - q + 1$, and 
$i^*$ has already a large voting power.\\

We start with parameter $\bar{\mathtt{req}}$.
\begin{theorem}\label{thrm: XP for req bar}
	\textbf{WMaxP} is in XP with respect to parameter $\bar{\mathtt{req}}$.
\end{theorem}
To prove this theorem we first show that \textbf{WMaxP} can be solved in polynomial time if $\bar{\mathtt{req}} = 0$. In this case, the problem can be solved by computing a directed spanning tree rooted in $i^*$ maximizing some weight function which can be achieved efficiently using Edmond's algorithm~\cite{edmonds1967optimum}.% (see  %Section~\ref{app : hardness2} in the Supplementary Material file).  

\begin{restatable}{lemma}{WTPReqBarZero}\label{lemma : WTP req bar is 0}
\textbf{WMaxP} can be solved in polynomial time if $\bar{\mathtt{req}} = 0$. 
\end{restatable}

Using Lemma~\ref{lemma : WTP req bar is 0}, we prove Theorem~\ref{thrm: XP for req bar}.
\begin{proof}[Proof of Theorem \ref{thrm: XP for req bar}]
As voters' weights are positive integers, the maximum number of voters that $i^*$ does not necessarily need the support of to reach the threshold is bounded by $\bar{\mathtt{req}}$. One can hence guess the set $C$ of voters that are not required with $|C|\le \bar{\mathtt{req}}$. Indeed, the number of possible guesses is bounded by $\frac{|V|^{\bar{\mathtt{req}}+1}- |V|}{|V| - 1}$. Let $X\subset V$ be one such guess. Once these voters are removed from the instance, we obtain another instance of \textbf{WMaxP} in which (if the guess is correct) $i^*$ should obtain the support of all other voters. This amounts to solving an instance of \textbf{WMaxP} where $\bar{\mathtt{req}} = 0$. Hence, it can be solved in polynomial time by Lemma~\ref{lemma : WTP req bar is 0}.
\end{proof}

Hence, interestingly \textbf{WMaxP} can be solved in polynomial time if $\bar{\mathtt{req}}$ is bounded by a constant. Unfortunately, \textbf{WMaxP} is W[1]-hard w.r.t. $\bar{\mathtt{req}}$ and hence is unlikely to be FPT for this parameter. 

\begin{restatable}{theorem}{WMaxPIsWHardReqBar}
	\textbf{WMaxP} is W[1]-hard with respect to $\bar{\mathtt{req}}$, even when all voters have weight one.
\end{restatable}
\begin{proof}
We design a parameterized reduction from the independent set problem. In the independent set problem, we are given a graph $G = (V,E)$ and an integer $k$ and we are asked if there exists an independent set of size $k$. The independent set problem is W[1]-hard parameterized by $k$. From an instance $I = (G= (V,E),k)$ of the independence set problem, we create the following \textbf{WMaxP} instance. We create a digraph $D = (\bar V, \bar A)$ where: 
\begin{itemize}
    \item $\bar V = U \cup W \cup \{i^*\} $ with $U = \{u_v : v\in V\}$ and $W = \{w_e, w_e^1,\ldots, w_e^{k} : e \in E\}$.
    \item $\bar A = \{(u_v,i^*) : v\in V\} \cup \{(w_e,u_v) : e\in E, v\in V, v \in e\} \!\cup\! \{(w_e^1,w_e),\ldots,$ $(w_e^k,w_e) : e \in E\}$.
\end{itemize}
All voters have weight one. The initial delegation function is such that $d(x) = x$ for $x\in U \cup \{i^*\}\cup\{w_e : e \in E\}$ and  $d(w_e^j) = w_e$ for each $j\in [k]$ and $e \in E$. The budget $\bar k = |E|+|V| - k$ and $\tau$ is set to $(k+1)|E| + |V| - k + 1$. Hence, $\bar{\mathtt{req}} = k$. We show that the instance of the independent set problem is a yes instance iff the instance of the \textbf{WMaxP} problem is a yes instance. To reach the threshold of $\tau$, $i^*$ necessarily needs the delegations of all voters $w_e$. This requires spending a budget of $|E|$ to make all voters $w_e$ delegate to some voters in $U$ (which should then delegate to $i^*$). Then, there only remains a budget $|V|-k$ to make these voters in $U$ delegate to $i^*$. Hence, we can reach the threshold $\tau$ iff we can make all voters in $\{w_e : e \in E\}$ delegate to less than $|V|-k$ voters in $U$. This is possible iff $I$ is a yes instance.        
\end{proof}

%Let us now consider parameter $\mathtt{req}$. 
Interestingly, \textbf{WMaxP} is FPT with respect to $\mathtt{req}$. 
\begin{restatable}{theorem}{WMaxPFPT} \label{thrm: FPT for req}
	\textbf{WMaxP} is FPT with respect to $\mathtt{req}$.
\end{restatable}
\begin{proof}[Proof sketch] %The complete proof can be found in the Supplementary Material file. %Appendix~\ref{app : hardness2}. 
Let $I = (\langle D = (V,A),\omega, d \rangle, i^*, k, \tau)$ be an instance of \textbf{WMaxP}. 
As voters' weights are positive integers, the maximum number of additional voters that $i^*$ needs the support of to reach the threshold is bounded by $\mathtt{req}$. 
We first note that one can collapse the tree $T_d(i^*)$ in one vertex with weight $\alpha_d(i^*)$.  
Let us consider a delegation function $d'$ such that $|\{i : d(i)\neq d'(i)\}| \le k$, and $\alpha_{d'}(i^*) \ge \tau$ (assuming such a solution exists). 
A subtree of $T_{d'}(i^*)$ rooted in $i^*$ with at most $\mathtt{req}+1$ voters accumulates a voting weight greater than or equal to $\tau$. 
Our FPT algorithm guesses the shape of this tree and  then looks for this tree in $D$ by adapting the color coding technique~\cite{alon1995color}. 
The idea is to color the graph randomly with $\mathtt{req}+1$ colors. 
If the tree that we are looking for is present in graph $D$, it will be colored with the $\mathtt{req}+1$ colors (i.e., one color per vertex) with some probability only dependent of $\mathtt{req}$. We say that such a tree is colorful. 
One can then resort to dynamic programming to find the best colorful tree rooted in $i^*$ in $D$ and which contains at most $k$ arcs not in $H_d$. 
This algorithm can then be derandomized using families of perfect hash functions~\cite{alon1995color,schmidt1990spatial}.

\end{proof}

\section{Power Distribution}

The structure of a delegation graph has a considerable impact on the power distribution among agents. In this section, we aim to find a delegation graph with a fixed number of gurus to maximize the minimum power an agent can have. 
Such delegation graph could for instance be useful to determine a committee within the set of voters in such a way that all voters feel important. 
Ensuring that all voters have a ``high'' relative importance incentives participation and prevents some voters to accumulate too much power.

More formally, given a digraph $D$, and a delegation graph $H \in \Delta(D)$, we denote by $f^{H}_i$ the power index of voter $i$ in $H$. The value of $f^{H}_i$ may be computed using the delegative Banzhaf index or the delegative Shapley-Shubik-index. Hence, the weakest agents in $H$ have a power index worth:

$$ \text{(weakest)}\ \ \mu(H):=\min_{i\in V} f^{H}_i.$$
%More formally, suppose we are give a digraph $D$ and a non-negative integer $k$. Consider a delegation graph $H \in \Delta(D)$ with $k$ gurus. Let $f^{H}_i$ be the power index of voter $i$ in $H$, the weakest agent in $H$ are:

The goal here is to organize the delegations in $H$ to maximize $\mu(H)$. We term this problem the MaxiMin Weight Problem (\textbf{MMWP} for short). Formally,

\begin{center}

\noindent\fbox{\parbox{14cm}{
		\textbf{MMWP}\\
		\emph{Input}: A graph $D$, a weight function $\omega$, a quota $q$ and an integer $k$.\\
		\emph{Feasible Solution}: A delegation graph $H = (V, E)\in \Delta(D)$ with $k$ gurus. \\
		\emph{Measure for \textbf{MWP}}: $\mu(H)$ to maximize, where $\mu(H)$ can be computed using either the delegative Banzhaf value or the delegative Shapley-Shubik value. 
}}\\
\end{center}

Consider the following observation from Zang and Grossi~\cite{zhang2020power},

\begin{fact}[Fact 2 in~\cite{zhang2020power}]\label{factMonotonocity}
For any pair of agents $i, j \in V(D)$, such that $d_i=j$, $DB_i \le DB_j$.
\end{fact}

Let us assume that $f^{H}_i$ is computed using the delegative Banzhaf value. 
Hence, using Fact~\ref{factMonotonocity}, it is worth noting that to give an optimal solution to \textbf{MMWP}, we only need to find a delegation graph $H_d$ in which $\min_{l \in Leaves(H_d)}f_l^{H_d}$ is maximized where $Leaves(H_d)$ is the set of all leaves in $H_d$. In the following, we will show that \textbf{MMWP} is NP-hard. 

\begin{theorem}\label{thMWP}
Problem \textbf{MMWP} is NP-hard.
\end{theorem}

To prove Theorem~\ref{thMWP}, we will give a reduction from the NP-complete $3$D-Matching problem~\cite{kleinberg2006algorithmdesign}. 
This problem is defined as follows. Let $X$, $Y$, and $Z$ be finite, disjoint sets, and $T \subseteq X \times Y \times Z$ be a set of ordered triples. That is, $T$ consists of triples $(x, y, z)$ such that $x \in X$, $y \in Y$, and $z \in Z$. Now $M \subseteq T$ is a $3$D-Matching if the following holds: for any two distinct triples $(x_1, y_1, z_1) \in M$ and $(x_2, y_2, z_2) \in M$, we have $x_1 \ne x_2$, $y_1 \ne y_2$, and $z_1 \ne z_2$.  In an instance $I=(T,k)$ of $3$D-Matching, the goal is to find a $3$D-Matching $M \subseteq T$ of size $|M|=k$. The resulting decision problem is NP-complete even when $|X|=|Y|=|Z|=k$.

From an instance $I=(T,k)$ of $3$D-Matching, we create an instance $I'=(D,\omega, q, k')$ of \textbf{MMWP} as follows. 
For any $x \in X$, $y \in Y$ and $z \in Z$, we create voters $v_x$, $v_y$ and $v_z$, respectively. 
We let $V_X := \{v_x : x \in X\}$, $V_Y := \{v_y : y \in Y\}$ and $V_Z = := \{v_z : z \in Z\}$ respectively. 
For any triple $(x,y,z) \in T$, we create two directed edges: one from $v_x \in V_X$ to $v_y \in V_Y$, and the other one from $v_y \in V_Y$ to $v_z \in V_Z$. We set the budget $k'$ to $k$ and the quota $q$ to $3$. All voters have voting weight 1. The power index which is used is the delegative Banzhaf index. 

We need the following observation to prove Theorem~\ref{thMWP}.

\begin{obs}\label{obsMWP}
Given an instance $I'=(D, \omega, q,k')$ resulting from our reduction, $\mu(H_d) \ge 2^{-k-1}$ iff for any subtree $T$ of $H_d$ rooted in a vertex in $V_Z$, $|T|=3$.
\begin{proof}
Since $k'=|Z|$, in a solution of \textbf{MMWP}, all voters $v_z \in V_Z$ should vote and all other voters $v_y \in V_Y$ and $v_x \in V_X$ should delegate to voters in $V_Z$ (directly or indirectly). Let $NS(v_x)$ be the set of coalitions $C$ that $v_x$ is a swing agent in. 
As $q=3$, any voter $v_x$ is a swing agent for a coalition $C$ iff:
\begin{itemize}
    \item $(c_d(v_x,d^*_{v_x})\setminus\{x\}) \subseteq C$, where $d^*_{v_x} \in V_Z$. ;
    \item the other voters of $C$ in $V_X \setminus c_d(v_x,d^*_{v_x})$ and $V_Y \setminus c_d(v_x,d^*_{v_x})$ are inactive agents in $C$;
    \item the coalition $C$ does not include another voter from $V_Z$ than $d^*_{v_x}$.
\end{itemize} 
Hence for any $v_x$, $|NS(v_x)| \le 2^{2k-2}$ as $NS(v_x) \subseteq \{(c_d(v_x,d^*_{v_x})\setminus\{x\}) \cup S : S \subseteq (V_Y\setminus\{d_{v_x}\})\cup (V_X\setminus\{v_x\})\}$. 
Moreover, for any $v_x$, $|NS(v_x)|=2^{2k-2}$ if $NS(v_x) = \{(c_d(v_x,d^*_{v_x})\setminus\{x\}) \cup S : S \in (V_Y\setminus\{d_{v_x}\})\cup (V_X\setminus\{v_x\})\}$. 
Stated otherwise, all voters in $V_X \setminus \{v_x\}$ and $V_Y \setminus \{d_{v_x}\}$ are always inactive agents in a coalition containing $c_d(v_x,d^*_{v_x})$ and no other voters in $V_Z$ than $d^*_{v_x}$.  
It is easy to realize that this condition is met in a delegation graph $H_d$ iff all subtrees rooted in a vertex of $V_Z$ in $H_d$ have size equal to $3$. This proves the obervation as the delegative Banzhaf index of a voter $v_x\in V_X$ is then equal to $|NS(v_x)|/2^{n-1} = 2^{-k-1}$  which proves the observation.
%Now imagine a delegation graph $H_d$ in which for any vertex $v_x$, all voters $(V_X\cup V_Y) \setminus c_d(v_x,d^*_{v_x})$ can join a coalition $S$ as inactive agent such that $c_d(v_x,d^*_{v_x}) \subseteq S$. The necessary condition for such a delegation graph $H_d$ is that 
\end{proof}
\end{obs}

We are now ready to prove Theorem~\ref{thMWP}.

\begin{proof}[Proof of Theorem~\ref{thMWP}]
Let $I=(T,k)$ be an instance of $3$D-Matching, and $I'=(D,\omega, q, k')$ be the instance of \textbf{MMWP} resulting from our reduction. 
We want to show that $I$ is a yes instance iff we can find a solution $H_d$ in $I'$ such that $\mu(H_d) \ge 2^{-k-1}$. 
For the first direction, suppose that we have a 3D-Matching $M \subseteq T$ of size $k=|X|=|Y|=|Z|$. 
Consider the delegation graph, where for any $(x,y,z) \in M$, $x$ delegates to $y$ and $y$ delegates to $z$, respectively. 
Hence, in the resulting delegation graph $H_d$, all subtrees rooted in a vertex of $V_Z$ are of size equal to $3$. Using Observation~\ref{obsMWP}, we obtain the desired conclusion. 
Let $H_d$ be a solution in $I'$ such that $\mu(H_d) \ge 2^{-k-1}$. Using Observation~\ref{obsMWP}, we obtain that all subtrees rooted in a vertex of $V_Z$ are of size equal to $3$. Notably, each of these subtrees necessarily contain one element of $V_X$, one element of $V_Y$ and one element of $V_Z$. There are $k$ of them and they do not share any vertex. Hence, they correspond to a 3D-Matching of size $k$. 
\end{proof}

%In the next step, we will show that \textbf{MSP} is NP-hard even when $k=1$. Note that for any $g \in Gu_d$, $f_g \ge f_v$, where $v \in T_d(g)$. Thus to give an optimal solution to \textbf{MSP}, it is enough to minimize $\max_{g\in Gu_d} f^{H_d}_g$. We will give a reduction from the Hamiltonian path problem to \textbf{MSP}.

%\begin{theorem}\label{thMSP}
%\textbf{MSP} is NP-hard even when $k=1$.

%\begin{proof}
%We prove the hardness via a reduction from the Hamiltonian path problem.Consider an instance $I=(V,E)$ of the Hamiltonian path problem, where $V$ is the set of voters and $E$ is the set of edges. To relate the Hamiltonian path problem and \textbf{MSP}, we need to create a graph $G'$ obtained from $I$ by adding a new voter $u$ and connecting all voters $V$ of $I$ to $u$. This means that $G'=(V',E')$, where $V'=V \cup \{u\}$ and $E'=E \cup \{(v,u)| v\in V\}$. Since $k=1$, obviously the only choice we have is to select $u$ as the guru for all voters. Let $H_d \in \Delta(G')$ be a delegation graph such that $u$ is the only guru and it only contains a path, starting from an arbitrary voter $v \in V'\setminus \{u\}$, passing all voters once and ending in $u$. We denote by $\mathcal{J}(G') \subset \Delta(G')$ the set of all such delegation graphs. As $k=1$, we see that for any $H'_d \in \mathcal{J}(G')$, $u$ has the minimum possible power index. It is easy to see that $I$ is a ``yes''-instance of the Hamiltonian path problem if and only $|\mathcal{J}(G')| \ge 1$. This implies that \textbf{MSP} is NP-hard.
%\end{proof}
%\end{theorem}
%\input{Problem/researchquestions}
\section{Conclusion}
Following a recent work by Zhang and Grossi~\cite{zhang2020power}, we investigated delegative simple games, a variant of weighted voting games in which agents’ weights are derived from a transitive support structure. We proposed a pseudo-polynomial time algorithm to compute the Banzhaf and Shapley-Shubik measures for this class of cooperative games and investigated several of their properties highlighting that they could lead to manipulations, e.g., by changing the delegation structure underlying the game. % and could characterize the delegative Shapley-Shubik index.
From this observation, we investigated a bribery problem in which we aim to maximize/minimize the power/weight of a given voter. We showed that these problems are NP-hard to solve and provided some more positive results (from the algorithmic viewpoint) by resorting to approximation algorithms and parameterized complexity. % destructive problem where one attempts to decrease the voter's power is NP-hard even when the weights of all voters are identical to one. In the constructive version, we proved that maximizing the agent's power is NP-hard even when voters' weights and the quota are given in unary. These hardness results comes from the fact that an agent's power does not only depend on the voting weight distribution, but also on the structure of the supports she receives from her followers. So, we considered a seemingly easier version of the constructive problem, where the goal is to maximize the voting weight of the agent. For this problem, we were able to show that designing an approximation algorithm with approximation ratio better than $1-1/e$ is not possible, unless P=NP and we also provided some parameterized complexity results. %Lastly, we proved that finding a delegation graph with a certain number of gurus that maximizes the minimum power an agent can have is an NP-hard problem.

Several directions of future work are conceivable. First, for both destructive and constructive bribery problems, designing some algorithms with tighter approximation guarantees under some conditions is one direction. Second, it would be interesting to study bribery problems related to alternative, maybe finer power measures. For instance, it is known that the Banzhaf index can be decomposed into two parts (the Coleman measures), one that measures the ability to initiate action, and one other to prevent it~\cite{felsenthal1998measurement}.

%it would be interesting to further investigate manipulation problems by splitting an agent in two and then sharing her delegations between the two resulting agents, a type of manipulation which was suggested by proposition~\ref{obs:md}.  %It would be interesting to investigate problems related to distribution of power among agents as finding a delegation graph with a fixed number of gurus that maximizes the minimum power a guru has. 

%\bibliographystyle{unsrt}
\bibliographystyle{plain}
\bibliography{biblio.bib}

\clearpage

%\section{Appendix}

%\begin{strip}%
{
 \centering
 \Huge  Appendix \\[1.5em]
 %The Supplementary Material
 }
%\end{strip}

%\twocolumn
%\nobalance

\section{Deferred proofs from Section~\ref{section : properties}} \label{App : properties}

\thShBaSudoPolynomial*
\begin{proof}
Consider a coalition $C \subseteq V$. Observe first that when a coalition $C$ does not include voter $j$, a subset $S$ of $j$'s followers (i.e., $S \subseteq T_d(j) \setminus \{j\}$) could join $C$ only as inactive agents and are not able to increase the total weight of $C$. For instance in Example~\ref{exNotations}, in the case voter $7$ does not join a coalition $C$ (i.e., $\gamma_{d,C}(8)=1$), all of her followers could join $C$ as inactive agents.

By relabeling the vertices we can make the assumption that the vertex we consider is voter $n$.  We first show how to compute the delegative shapely value.
We decompose the proof in two sub-cases :

\paragraph{Case 1: voter $n$ is a guru.} 
We run a depth-first search of $H_d$ starting from voter $n$. 
Let $\pi_d$ be the ordering of voters in $H_d$ such that voters are sorted in decreasing order of their discovering times in the depth first search. 
As $n$ is first visited in the depth first search, it is the last voter in the ordering $\pi_d$. In Example~\ref{exNotations}, voters are sorted in such a way (i.e., $\pi_d=<1, 2, 3, 4, 5, 6, 7, 8>$), when computing $DB_8(\mathcal{E})$. 
Consider a guru $n$ with accumulated weight $\alpha_d(n)$. %For every $k \in \{0, ..., \alpha_d(n)-w_n\}$, let $T_d(n,k)$ be the set of all subtrees $T$ rooted in $n$ such that $\sum_{i \in T}w_i=k+w_n$, i.e., each subtree $T \in T_d(n,k)$ represents a possible way that $n$ obtains an accumulated weight $k+w_n$. 
%It is worth noticing that there may be more than one subtree $T \in T_d(n,k)$ in which voter $n$ receives weight $k+w_n$. For instance, in Example~\ref{exNotations}, $T_d(8)$ has two subtrees with accumulated weight $3$ and rooted in $8$ : the first (resp. second) subtree is composed of vertices in $\{6,7,8\}$ (resp. $\{4,7,8\}$). Given a subtree $T$, let $ch(T)$ be the set of children of nodes in $T$. To compute $DS_n$, we need to compute, for every $T \in T_d(n,k)$, $N^k_s(T)$ the number of $s$-element subsets of $V\setminus (T \cup ch(T))$ that have weight in $\{q-k-w_n, ..., q-1\}$, where $s=0, 1, ..., n-|T|-1$ and $k=0, ..., \alpha_n(d)-w_n$. 
Recall that $w_n$ is the initial weight that voter $n$ is endowed with before starting the delegation process and $\alpha_d(n)$ is the total weight that voter $n$ receives through delegations in $H_d$. 
%More formally,  for every $T \in T_d(n,k)$, we have 
%$$N^k_s(T)=\sum_{\substack{C\subseteq V\setminus (T \cup ch(T))\\ 
%|C|=s}} \mathbbm{1}_{q-k-w_n \le \alpha_{d}(C) \le q-1},$$

%where $\mathbbm{1}_{\phi}$ is equal to $1$ if $\phi$ is true, and $0$ otherwise. 
%Let $N_s = \sum^{ \alpha_n(d)-w_n }_{k = 0}\sum_{T \in T_d(n,k)} N^k_{s}(T)$. 
Let $N_s$ denote the number of coalitions $C$ of size $s$ for which $n$ is a swing agent. 
%Now we can compute $DS_n$ as follows. 
Then $DS_n$ can be rewritten as follows.
\begin{align}\label{alignShapely}
    DS_n = \sum_{s=0}^{n-1}\frac{s!(n-s-1)!}{n!}N_s. 
\end{align}

Let $t_j=|T_d(j)|$, i.e., the size of the tree $T_d(j)$ rooted in a vertex $j$. 
Moreover, let $Del(j) = c_d(j,d^*(j))\setminus\{j\}$ be the set of voters to which $j$ delegates to directly or indirectly (not to be confused with the guru of $j$).
To compute $DS_n$ efficiently, we use a dynamic programming approach. 
Given a forest $\mathcal{F}$ with $n$ vertices, we define a table $F_{\mathcal{F}}$ where $F_{\mathcal{F}}[j, w, s]$ is the number of sets $C\subset \{1,...,j\}$ such that $|C| = s$ and $\gamma_d(Del(j)\cup C) = \gamma_d(Del(j)) + w$, where $j$ and $s$ range from $0$ to $n$ and $w$ ranges from $0$ to $\sum_{i \in V}w_i$. In words, we count the number of s-element subsets in $\{1,...,j\}$ who will contribute a weight $w$ when added to a coalition containing all elements of $Del(j)$ (this ensures that $j$ will be an active agent). These table entries will be filled as follows.

First, for $j=0$, we have
\begin{align*}
F_{\mathcal{F}}[0, w, s] \!=\! \left\{ \begin{array}{c c}
1 & \text{ if } w= 0 \text{ and } s=0\\
0 & \text{ otherwise.}
\end{array} \right.
\end{align*}
More generally, $F_{\mathcal{F}}[j, w, s] = 0$ if $s > j$ or $w > \sum_{l=1}^j w_{l}$. Second, if $s = 0$, then $F_{\mathcal{F}}[j, w, s] = 0$ if $w > 0$ and $1$ otherwise. Third, in the special case $w=0$, and $j\ge s > 0$, we have
\begin{align}\label{alignWZeroVoters}
         F_{\mathcal{F}}[j, 0, s]&=\sum_{x = 0}^{ t_j-1} \binom{t_j-1}{x} F_{\mathcal{F}}[j - t_j, 0, s - x].
\end{align}
Indeed, in this case $j$ cannot be part of the coalition as she would contribute a positive weight. Conversely, as $j$ is not part of the coalition, any subset of vertices in $T_d(j) \setminus\{j\}$ may be part of the coalition as a set of inactive agents. We consider each possible cardinality (variable $x$) for such a subset before moving to the next tree rooted in $j-t_j$.

Now for $j=1,\ldots, n$, $s=1,\ldots, j$, and $w=1, \ldots, q-w_n$, $F[j, w, s]$ will be computed  as follows.

\begin{equation}\label{eqRecursiveF}
      \begin{split}
         F_{\mathcal{F}}[j, w, s]&=\sum_{x = 0}^{ t_j-1} \binom{t_j - 1}{x} F_{\mathcal{F}}[j - t_j, w, s - x]\\&+ F_{\mathcal{F}}[j-1,w-w_j, s-1]
      \end{split}
\end{equation}
We recall that $F_{\mathcal{F}}[j, w, s]$ is the number of sets $C\subset \{1,...,j\}$ such that $|C| = s$ and $\gamma_d(Del(j)\cup C) = \gamma_d(Del(j)) + w$. Here, the two summands correspond to the cases $j\notin C$ and $j\in C$ respectively.  Note that in the first case, voters in $T_d(j)$ can only join $C$ as inactive agents. Hence in the first summand, we consider all possible ways of adding to $C$ inactive agents from the tree rooted in $j$ before moving to the next subtree (rooted in $j - t_j$).

We use this dynamic programming equations considering two forests $\mathcal{F}_1 = \{T_d(n)\}$ and $\mathcal{F}_2 = H_d - T_d(n)$ of sizes $t_n$ and $n-t_n$, respectively. We conclude by noticing that:
\begin{equation}
N_s = \sum_{k = 0}^s \sum_{w = 0}^{q-1} F_{\mathcal{F}_2}[n-t_n,w,k] (\sum_{w' = q - w}^{\alpha_d(n)} F_{\mathcal{F}_1}[t_n,w',s-k+1]).\label{eq:Ns computation}
\end{equation}

\paragraph{Case 2: voter $n$ is a delegator.}
If $n$ is a delegator, then $n$ can only be a swing agent for coalition $C$ if $Del(n) \subseteq C$. 
Hence, we should count only coalitions which meet this condition. 
To take this fact into account we use our dynamic programming approach on the delegation graph $H_d[V\setminus Del(n)]$ (the new roots are now considered as voting agents). 
Moreover, in this case $N_s$ is obtained by a different formula. Let $C_\delta = Del(n)$, then  we use the formula:   
\begin{equation} \label{eq:Ns computation2}
\begin{multlined}
N_s = \sum_{k = 0}^{s_\delta} \sum_{w = 0}^{q_\delta-1} F_{\mathcal{F}_2}[n-t_n,w,k] \times \\(\sum_{w' = q_{\delta} - w}^{\alpha_d(n)} F_{\mathcal{F}_1}[t_n,w',s_{\delta}-k+1]),
\end{multlined}
\end{equation}
where $s_\delta = s - |C_\delta|$ and $q_\delta = q - \gamma_d(C_\delta)$.

We move to the complexity analysis. %All values $(l \choose k)$ and $k!$ (for $0\le k \le l\le n$) can be computed in $O(n^2)$ and $O(n)$ respectively by dynamic programming. 
The running time of this algorithm is bounded by the time required to fill out the tables $F_{\mathcal{F}_1}$ and $F_{\mathcal{F}_2}$. The size of each table is bounded by $n \times nw_{\max} \times n$ and Equation \ref{eqRecursiveF} can be computed in $O(n)$ operations (if values $\binom{n}{k}$ are precomputed). This leads to a complexity of $O(n^4w_{\max})$ for computing tables $F_{\mathcal{F}_1}$ and $F_{\mathcal{F}_2}$. As Equations \ref{eq:Ns computation} and \ref{eq:Ns computation2} can be computed in $O(n^3w_{\max}^2)$, %$O(n^3)$
computing all values $N_s$ can be performed with $O(n^4w_{\max}^2)$ operations. Last, we compute the Shapley value which requires $O(n)$ operations. 

For the Banzhaf index, we can simply adapt the dynamic program and omit the third index $s$. Thus, the complexity to fill out both adapted tables $F_{\mathcal{F}_1}$ and $F_{\mathcal{F}_2}$ in this case will be upper bounded by $O(n^2w_{\max})$, as the size of each table is bounded by $n^2w_{\max}$ and each cell can be filled out by $O(1)$ operations. To compute the Banzhaf index of each agent $n$, we need to find the number of subsets $N\setminus T_d(n)$ having weight at least $q-\alpha_d(n)$ and at most $q-1$. This can be done in $O(n^2w_{\max}^2)$ times as both $q$ and $\alpha_d(n)$ are bounded by $nw_{\max}$. This leads to a complexity of $O(n^2w_{\max}^2)$ for computing the Banzhaf index.
\end{proof}

We now investigate an additional axiom compared to what is provided in the main text. We borrow the following definition from Zhang and Grossi~\cite{zhang2020power}. 
%\begin{definition}[Unanimity]
%$\mathcal{E}$ is a unanimity LDE if $q=\sum_{i \in V}\omega(i)$, denoted by $\mathcal{UE}$.
%\end{definition}

%Thus, in a unanimity delegative simple game $\mathcal{G}_{\mathcal{UE}}$, a coalition $C$ wins if $\alpha_d(C) \ge \sum_{i \in V}\omega(i)$.

\begin{definition}[Composition]\label{defComposition}
Let two LDEs $\mathcal{E}_1 = \langle D_1 = (V_1 , A_1), \omega_1, d^1, q_1 \rangle$ and $\mathcal{E}_2 = \langle D_2 = (V_2 , A_2), \omega_2, d^2, q_2 \rangle$, such that for any $i \in V_1\cap V_2$, if $d^1(i)=j$ (resp. $d^2(i)=j$), then $j\in V_2$ (resp. $j \in V_1$) and $d^2(i)=j$ (resp. $d^1(i)=j$), and $\omega_1(i)=\omega_2(i)$. We define two new delegative simple games $\mathcal{E}_1 \land \mathcal{E}_2=\langle D_{1\land 2}(V_1 \cup V_2, A_1 \cup A_2), \omega_{1\land 2}, d_{1\land 2}, q_1 \land q_2 \rangle$ and $\mathcal{E}_1 \lor \mathcal{E}_2=\langle D_{1\lor 2}(V_1 \cup V_2, A_1 \cup A_2), \omega_{1\lor 2}, d_{1\lor 2}, q_1 \lor q_2 \rangle$, where:

\begin{itemize}
    \item for any $i \in V_1$ (resp. $i \in V_2$), $\omega_{1\lor 2}(i)=\omega_{1 \land 2}(i)=\omega_1(i)$ (resp. $\omega_{1\lor 2}(i)=w_{1 \land 2}(i)=\omega_2(i)$);
    
    \item for any $i \in V_1 \setminus V_2$ (resp. $V_2\setminus V_1$), $d_{1\lor 2}(i)=d_{1 \land 2}(i)=d^1(i)$ (resp. $d_{1\lor 2}(i)=d_{1 \land 2}(i)=d^2(i)$), and for any $i \in V_1 \cap V_2$, as $d^1(i)=d^2(i)$ by initial conditions, then $d_{1\lor 2}(i)=d_{1 \land 2}(i)=d^1(i)=d^2(i)$;%, otherwise $d_{1\lor 2}(i)=d_{1 \land 2}(i)=d^k(i)$ where $k \in \{1,2\}$ and $d^k(i)\ne 0$;
    
    \item $q_1\land q_2$ (resp. $q_1\lor q_2$) is met iff $\sum_{i\in C \cap V_1} \gamma_{d,C\cap V_1}(i) \ge q_1$ and (resp. or)\\ $\sum_{i\in C \cap V_2} \gamma_{d,C\cap V_2}(i)\ge q_2$.

\end{itemize}
\end{definition}

Put another way, when two LDEs $\mathcal{E}_1$ and $\mathcal{E}_2$ coincide in their intersection, by composing $\mathcal{E}_1$ and $\mathcal{E}_2$, we obtain two new LDEs $\mathcal{E}_1 \land \mathcal{E}_1$ and $\mathcal{E}_1 \lor \mathcal{E}_2$. Let $\mathcal{G}_{\mathcal{E}_1 \lor \mathcal{E}_2}$ (resp. $\mathcal{G}_{\mathcal{E}_1 \land \mathcal{E}_2}$) be the delegative simple game induced by $\mathcal{E}_1 \lor \mathcal{E}_2$ (resp. $\mathcal{E}_1 \land \mathcal{E}_1$). In $\mathcal{G}_{\mathcal{E}_1 \lor \mathcal{E}_2}$ (resp. $\mathcal{G}_{\mathcal{E}_1 \land \mathcal{E}_2}$), a coalition $C \subseteq V_1 \cup V_2$ wins if $C$ wins in $\mathcal{G}_{\mathcal{E}_1}$ or (resp. and) $\mathcal{G}_{\mathcal{E}_2}$. Note that according to our setting, Definition~\ref{defComposition} does not consider the case where voters abstain, which is the only difference between Definition~\ref{defComposition} and the one given by Zhang and Grossi~\cite{zhang2020power}. 

We now investigate an additional axiom from Zhang and Grossi~\cite{zhang2020power}. 

\begin{axiom}[Sum Principle (\textbf{SP})]\label{axiomSP}
For any two LDEs ${\mathcal{E}}_1$ and ${\mathcal{E}}_2$, such that any $i \in V_1 \cup V_2$ satisfies the condition in Definition~\ref{defComposition}, $f_i(\mathcal{E}_1 \land \mathcal{E}_2)+f_i(\mathcal{E}_1 \lor \mathcal{E}_2)=f_i(\mathcal{E}_1)+f_i(\mathcal{E}_2)$.
\end{axiom}

Zhang and Grossi~\cite{zhang2020power} showed that the delegative Banzhaf index satisfies axiom $SP$. We show that it is also the case of the delegative Shapley-Shubik index. 
\begin{proposition}
The delegative Shapley-Shubik index satisfies axiom %\textbf{DP}, \textbf{EI}, \textbf{SA}, 
\textbf{SP}.
\end{proposition}
\begin{proof}

To show that $DS$ satisfies axiom \textbf{SP}, we recall that for any $C \subseteq V_1 \cup V_2$, $\nu_{\mathcal{E}_1 \land \mathcal{E}_2}(C)=1$ (resp. $\nu_{\mathcal{E}_1 \lor \mathcal{E}_2}(C)=1$) iff $\nu_{\mathcal{E}_1}(C \cap V_1)=1$ and (resp. or) $\nu_{\mathcal{E}_2}(C \cap V_2)=1$. 
Let $P_i^{\mathcal{E}_j}$ (with $j\in \{1,2\}$) be the set of permutations for which agent $i$ is a swing agent in $\mathcal{G}_{\mathcal{E}_j}$, i.e., $P_i^{\mathcal{E}_j}=\{\pi \in \Pi_j: m^{\nu_{\mathcal{E}_j}}_{\pi}(i) = 1\}$, %\nu_{\mathcal{E}_j}(C_{\pi}(i))= 0\text{ and } \nu_{\mathcal{E}_j}(C_{\pi}(i) \cup \{i\})=1\}$, 
where $\Pi_j$ is the set of all permutations of $V_j$. 
Let $n_1=|V_1|$, $n_2=|V_2|$ and $n = |V_1 \cup V_2|$. Consider a permutation $\pi$ of $V_j$ ($j \in \{1,2\}$). Let $\pi'_\pi$ be a permutation of $V_1 \cup V_2$ in which voters in $V_j$ has the same order as in $\pi$. The number of such permutations $\pi'_\pi$ is $\frac{n!}{n_j!}$.

Given a permutation $\pi$ of $V_1 \cup V_2$, we will denote by $\pi[V_1]$ (resp. $\pi[V_2]$) the restriction of $\pi$ to $V_1$ (resp. $V_2$). We will now count the permutations for which $i$ is a swing in $ \mathcal{G}_{\mathcal{E}_1 \lor \mathcal{E}_2}$. 
Note that a necessary condition for $i$ to be swing in $\pi$ for $\mathcal{G}_{\mathcal{E}_1 \lor \mathcal{E}_2}$ or $\mathcal{G}_{\mathcal{E}_1 \land \mathcal{E}_2}$ is that $i$ should be a swing in $\pi[V_1]$ for $\mathcal{G}_{\mathcal{E}_1}$ or $i$ should be a swing in $\pi[V_2]$ for $\mathcal{G}_{\mathcal{E}_2}$.

Given a permutation $\pi \in V_1$, for which $i$ is a swing agent in $\pi$ for $\mathcal{G}_{\mathcal{E}_1}$, then $\pi$ can be extended in $\frac{n!}{n_1}$ permutations $\pi'_\pi$ such that $\nu_{\mathcal{E}_1}(C_{\pi'_\pi}(i) \cup \{i\} \cap V_1) = 1 $ and $\nu_{\mathcal{E}_1}(C_{\pi'_\pi}(i) \cap V_1) = 0 $. 
For such a permutation $\pi'_\pi$, $i$ will be swing for $\mathcal{G}_{\mathcal{E}_1 \lor \mathcal{E}_2}$ if $\nu_{\mathcal{E}_2}(C_{\pi'_\pi}(i) \cap V_2) = 0 $. 
Moreover $i$ will then be a swing in $\pi'_\pi$ for $\mathcal{G}_{\mathcal{E}_1 \land \mathcal{E}_2}$ if $\nu_{\mathcal{E}_2}(C_{\pi'_\pi}(i) \cap V_2) = 1 $ or if $i$ is also a swing in $\pi'_{\pi}[V_2]$ for $\mathcal{G}_{\mathcal{E}_2}$. Hence, note that if $i$ is not a swing for $\pi'_\pi$ in $\mathcal{G}_{\mathcal{E}_1 \lor \mathcal{E}_2}$, then $i$ is a swing for $\pi'_\pi$ in $\mathcal{G}_{\mathcal{E}_1 \land \mathcal{E}_2}$.

A symmetric analysis can be  applied for permutations $\pi \in V_2$, for which $i$ is a swing agent in $\pi$ for $\mathcal{G}_{\mathcal{E}_2}$. From this analysis, we can derive the following formula:
\begin{equation}
        |P_{i}^{\mathcal{E}_1 \lor \mathcal{E}_2} |= \frac{n!}{n_1!}|P_i^{\mathcal{E}_1}| + \frac{n!}{n_2!} |P_i^{\mathcal{E}_2}| - |P_{i}^{\mathcal{E}_1 \land \mathcal{E}_2}|. \label{alignSPGeneral}
\end{equation}
As a last comment to justify the formula, note that if a permutation $\pi \in \Pi[n]$ is counted twice in the two first summands because $i$ is both a swing in $\pi[V_1]$ for $\mathcal{G}_{\mathcal{E}_1}$ and a swing in $\pi[V_2]$ for $\mathcal{G}_{\mathcal{E}_2}$, then this is corrected by the fact that $i$ is also a swing in $\pi$ for $\mathcal{G}_{\mathcal{E}_1 \land \mathcal{E}_2}$.

Now if we divide each side of Equation~\ref{alignSPGeneral} by $n!$, then we have 
\begin{equation*}
    \frac{|P_i^{\mathcal{E}_1 \lor \mathcal{E}_2}|}{n!}=\frac{|P_i^{\mathcal{E}_1}|}{n_1!} + \frac{|P_i^{\mathcal{E}_2}|}{n_2!}-\frac{|P_i^{\mathcal{E}_1 \land \mathcal{E}_2}|}{n!}
\end{equation*}
This implies that for any $i \in V_1 \cup V_2$, we have $DS_i(\mathcal{E}_1 \lor \mathcal{E}_2)+DS_i(\mathcal{E}_1 \land \mathcal{E}_2)=DS_i(\mathcal{E}_1)+DS_i(\mathcal{E}_2)$ and the proof is complete.

\end{proof}

\section{Deferred proofs from Section~\ref{section : PIM}} \label{app : hardness}

Before giving the proof of Theorem~\ref{thrm : IMinP is NP-Hard}, we introduce the concept of a distant agent. In delegation graph $H_d$, a voter $i$ is called a \textit{distant} agent if $\gamma_d(c_d(i, d^*_i) \setminus \{i\}) \ge q$.

\IMinPIsHard*
\begin{proof}
The membership in NP is implied by Theorem~\ref{thShBaPolynomial}. 
For the hardness part, we use a reduction from the NP-complete Hamiltonian path problem~\cite{garey1979computers}. Recall that in the Hamiltonian path problem, the goal is to determine if there exists a path in an undirected graph that visits each vertex exactly once.  
Consider an instance $I = (V,E)$ of the Hamiltonian path problem, where $V$ is the set of vertices and $E$ is the set of edges.  
We create the following instance from $I$ which is the same for both \textbf{SMinP} and \textbf{BMinP}. We create a digraph $D = (\bar V, \bar E)$ where $\bar V = V\cup\{u\}$ is obtained from $V$ by adding an additional vertex and $\bar E = \{(i,j),(j,i) : \{i,j\} \in E \} \cup \{ (u,i) : i \in V\}$ is obtained from $E$ by adding an arc from $u$ to every other vertex (and making the graph directed). Now consider the delegation graph $H_d \in \Delta(D)$, where all voters vote directly, i.e., $\forall i\in \bar V, d(i) = i$. We set $w_i=1$, for every voter $v_i \in \bar V$. Also, we set $k=|V|$, $q=|V|$ and $\mathcal{E} = \langle D, \omega, d , q \rangle$. Suppose that the goal is to modify $d$ (yielding another delegation function $d'$) in order to make $u$ a distant agent. Indeed, note that $DB_u(\mathcal{G}_{\mathcal{E}'}) = 0$ (resp. $DS_u(\mathcal{G}_{\mathcal{E}'}) = 0$) iff $u$ is a distant agent. We can make $u$ a distant agent in $\mathcal{E}'$ iff there exists a Hamiltonian path in $I$. Indeed, if there exists an Hamiltonian path $(v_1,v_2,\ldots, v_n)$ in $G$ we can make $u$ a distant agent by adding delegations $d'(v_i) = v_{i+1}, \forall i\in [n-1]$  and making $u$ delegate to $v_1$. Conversely, if we can make $u$ a distant agent, then $u$ is at the end of a simple path of length $|V|$ implying that there exists a Hamiltonian path in $I$.
\end{proof}

%%%%%%%%%%%%%%%
%CoNPHardness of IMinP On Trees

\IMinPIsCoHardOnTrees*
\begin{proof}
Consider a weighted voting game $\mathcal{G} = \langle V', \nu, q' \rangle$, where $V' = [n]$ is a set of $n$ agents. An agent $i$ is said to be a dummy player if for any coalition $C \subseteq V'\setminus\{i\}$, $\delta_i(C):=\nu(C\cup\{i\}) - \nu(C)=0$. In the Dummy Player problem, the goal is to find out that whether $i$ is a dummy player in $\mathcal{G}$. The Dummy Player problem is known to be a coNP-complete problem~\cite{chalkiadakis2016weighted}. We transform an instance $J=\langle V', \nu, q', i \rangle$ of the Dummy Player problem to an instance $I=\langle \mathcal{E} = \langle D= (V, A), \omega, d , q \rangle , v^*, k, \tau \rangle$ of \textbf{BMinP} (resp. \textbf{SMinP}).
We create a tree $D$ where: 
\begin{itemize}
    \item $V = V' \cup \{v\}$.
    \item $ A = \{(v_j, v): v_j \in V \setminus\{v\}\}$.
\end{itemize}

Let $R = \sum_{v' \in V'} w(v')$, where $w(v_j)$ is $v_j$'s weight in $J$. We set $q = R +q'$, $v^*=i$, $k=0$ and $\tau=0$ for both \textbf{BMinP} and \textbf{SMinP}, where $N=|V|$. The weight function $\omega$ is set as follows. $\omega(v')=w(v')$ for any $v' \in V'$ and $\omega(v)=R$. The initial delegation function $d$ is set as follows. $d(v_j)=v_j$ for any $v_j \in V$. %$k$ can be set from $1$ to $n=|V'|$.

Note that $v$ should be in any winning coalition $C'$ in $I$. Suppose that $J$ is a ``yes''-instance then $i$ is not a swing agent for any coalition $C \subseteq V'$. This implies that $v^*$ is not a swing agent for any coalition $C'=C \cup \{v\}$ in $I$. Conversely, suppose that $I$ is a ``yes''-instance, meaning that after changing at most $k$ delegations in $I$ there is no coalition $C'$ for which $v^*$ is a winning coalition, implying that no coalition $C=C'\setminus \{v\}$ in $J$ for which $i$ is a winning coalition. This concludes the proof.

%Now, assume, for a contradiction, that there is a factor $\alpha$ ($0 <\alpha \le 1$) polynomial time approximation algorithm, $\mathcal{A}$, for \textbf{BMinP} (resp. \textbf{SMinP}). Let $\mathcal{E}' = \langle D= (V, A), \omega, d' , q \rangle$, where $|\{i\in V : d(i)\neq d'(i)\}|\le k$, be the delegative simple game after applying $\mathcal{A}$. Thus 1) if $\mathcal{A}$ outputs $DB_{v^*}(\mathcal{E}')= 0$ (resp. $DS_{v^*}(\mathcal{E}') = 0$), then $J$ is a ``yes''-instance 2) if $\mathcal{A}$ outputs $DB_{v^*}(\mathcal{E}')> 0$ (resp. $DS_{v^*}(\mathcal{E}') > 0$), then $J$ is not a ``yes''-instance. We see that based on these two cases, the algorithm $\mathcal{A}$ can be used to decide whether $J$ is a ``yes''-instance or not, which is impossible, unless $P=coNP$. This concludes the proof.

\end{proof}
\IMaxPIsHard*
\begin{proof}
Membership in NP is induced by Theorem~\ref{thShBaPolynomial}. 
For the hardness part, we make a reduction from the NP-complete exact cover by 3-sets problem (X3C)~\cite{garey1979computers}. 
In the X3C problem, we are given a universe $U = \{x_1,\ldots, x_{3n}\}$ of $3n$ elements and a collection $T = \{S_1,\ldots, S_m\} \subset U^3 $ of $m$ subsets of $U$, each containing 3 elements.
The question asked is to determine if there exists a subcollection of $T$ which covers each element of $U$ exactly once.  

Given an instance $(U,T )$ of X3C, with $|U| = 3n$ and $|T| = m$, we create the following instance which is the same (up to the threshold $\tau$) for problems \textbf{BMaxP} and \textbf{SMaxP}. 
The voter set $V= \{v^*,v_d\} \cup V_U \cup V_T$ is composed of one voter $v_d$ with weight $N = 9n^2m + 9 n$, $3n$ element voters $V_U = \{v_{x} : x \in U\}$ with weight $\bar N = 3nm + 3 $, $3mn$ set voters $V_T = \{v_{S^{i}} : S \in T, i\in [3n]\}$ with weight 1, and one voter $v^*$ with weight 1. 
Remark that weights depend only polynomially on $n$.
The digraph $D = (V,A)$ has the following arcs. 
We have arcs from each element voter $v_x$ to each set voter $v_{S^{1}}$ for which $x \in S$, one arc from each set voter $v_{S^{i}}$ to set voter $v_{S^{i+1}}$ for $i\in[3n-1]$ and one arc from each set voter $v_{S^{3n}}$ to $v^*$. The initial delegation graph has no delegations, i.e., $\forall i\in V, d(i)=i$. 
The quota $q$ is set to $9n^2m+3nm+9n+2$. This value is chosen such that 1) $N+\bar N > q$, 2) $3n \bar N + 3nm + 1  < q$, and 3) $N+3nm+1 < q$. Lastly, the budget $k$ is set to $3n + 3n^2$. 

Let us first observe that a coalition $C$ will be successful iff it includes $v_d$ and at least one element voter as active voters. 
Hence, if $v^*$ is a swing agent for a coalition $C$ then first $v_d\in C$, second no element voter is active in $C$, and third $C$ should contain an element voter and a path from this element voter to $v^*$ through a path $(v_{S^1},v_{S^2}, \ldots,v_{S^{3n}},v^*)$ for some set $S\in T$. 

Let $H = (V,L)$ be a delegation graph which is a feasible solution in the \textbf{BMaxP} (resp. \textbf{SMaxP}) instance resulting from the reduction. 
We will say that the $S$-path is active in $H$, if $(v_{S^i},v_{S^{i+1}})\in L$ for $i\in [3n-1]$, and  $(v_{S^{3n}},v^*)\in L$. 
Note that the number of active paths in $H$ is at most $n$ (we do not consider the only feasible solution $H$ with $n+1$ active paths, as $v^*$ is never a swing agent for this solution). 
%For a coalition $C$, let $P(C) \subseteq T$ be the subset of sets $S\in T$ for which the $S$-path is active. 
For a set $S\in T$, we denote by $E_H(S)$ the set of element voters $v_x$ such that $(v_x, v_{S^1})\in L$.

Let us assume that there exists an exact cover by 3 sets $\{T_{i_1},\ldots, T_{i_n}\}$. 
Note that by relabeling the indices we can assume that $\{T_{i_1},\ldots, T_{i_n}\} = \{T_{1},\ldots, T_{n}\}$ and $T_i = \{x_{3(i-1)+1}, x_{3(i-1)+2}, x_{3i}\}$. 
We create the following delegation graph $H$. 
We make all voters $v_{x_i}$ delegate to $v_{T_{\lceil i/3 \rceil}^{1}}$ for $i\in[3n]$, all voters $v_{T_i^j}$ delegate to $v_{T_i^{j+1}}$ for $i\in[n]$ and $j\in [3n-1]$ and all voters $v_{T_i}^{3n}$ for $i \in [n]$ to $v^*$. 
Note that (up to the relabeling) $H$ does not depend on the specific exact cover by 3 set. However, such a solution $H$ is feasible only if the X3C instance is a yes instance. Let us show that such a solution $H$ achieves the highest possible delegative Banzhaf and Shapley-Shubik indices for agent $v^*$. Indeed, let us consider another feasible solution $H'$ and let $\{T_{i_1},\ldots,T_{i_t}\}$ be the sets in $T$ whose paths are active in $H'$. By relabeling, we can assume that $\{T_{i_1},\ldots,T_{i_t}\} \subseteq \{T_{1},\ldots,T_{n}\}$ and $E_{H'}(T_i)\subseteq \{x_{3(i-1)+1}, x_{3(i-1)+2}, x_{3i}\}$. The claim then follows by noticing that the set of coalitions for which $v^*$ is a swing agent in $H'$ is a subset of the set of coalitions for which $v^*$ is a swing agent in $H$ and is even a proper subset if $H \neq H'$ (after relabeling).   

We conclude by noticing that the delegative Banzhaf and Shapley-Shubik indices of a solution $H$ corresponding to an exact cover by 3-set can be easily computed. Indeed, this is simply implied by Theorem~\ref{thShBaPolynomial}.

\paragraph{Banzhaf value.} Given such a solution $H$, $v^*$ will be a swing agent for all coalitions that contain an element voter $x$ and an $S$-active path such that $x\in E_H(S)$. The number of such coalitions is
\begin{equation*}
    \sum_{k=1}^n \binom{n}{k} (2^{3n} - 1)^{n-k} 2^{3n(m-n)} (2^{3k}-1)2^{3(n-k)}.
\end{equation*}

Let us explain this formula. We count coalitions which include $k$ active-paths out of the $n$ active paths of $H$, and sum the obtained values for $k$ ranging from 1 to $n$. For a given $k$, these coalitions cannot include completely any of the $n-k$ other active paths. Hence, the number of possible intersections with the set of elements in these $n-k$ paths is of $(2^{3n} - 1)^{n-k}$. 
Conversely there is no constraint on set elements $v_{T_i^j}$ for $i\notin[n]$. Hence, the number of possible intersections with these set elements is worth $2^{3n(m-n)}$. The $k$ active paths in the coalition makes it possible to connect $3k$ element voters to $v^*$. As at least one of them should be included in the coalition, this leaves $(2^{3k}-1)$ possibilities. There are no constraints on the $3(n-k)$ other element voters explaining the last $2^{3(n-k)}$ term. This induces the Banzhaf value of voter $i^*$ for solution $H$. 

\paragraph{Shapley-Shubik value.} We introduce the Table $\Lambda$ where $\Lambda[i,j]$ counts the number of ways we can choose $i$ elements in a set of the form $\{v_{S_l^{t}}, l\in [j], t\in [3n]\}$ without taking all $3n$ elements $\{v_{S^{t}}, t\in [3n]\}$ of a same set $S$. Table $\Lambda$ can be computed efficiently using the following dynamic programming equations: 
\begin{align*}
    \Lambda[i,1]   &= \binom{3n}{i} \text{ if } i < 3n \text{ and } 0 \text{ otherwise}\\
    \Lambda[i,j>1] &= \sum\limits_{\substack{l=0\\l \le i}}^{3n-1} \binom{3n}{l} F[i-l,j-1].
\end{align*}
The Shapley value of voter $i^*$ for a solution $H$ corresponding to an exact cover by 3 set can then be derived by the following formula:
\begin{align*}
N_S =& \sum\limits_{\substack{k=1\\ 3nk \le S-2}}^n \binom{n}{k} \sum\limits_{\substack{i=1\\ 3nk+i \le S-1}}^{3k} \binom{3k}{i} \sum\limits_{\substack{j=0\\3nk+i+j\le S-1}}^{3(n-k)} \binom{3(n-k)}{j}\\
& \sum\limits_{\substack{l=0\\3nk+i+j+l\le S-1}}^{3n(m-n)}  \binom{3n(m-n)}{l} \Lambda[S-1-3nk-i-j-l,n-k]
\end{align*}
Let us explain this formula. We are counting the number of coalitions $C\subset V$ of size $S$ for which $i^*$ is a swing voter. A necessary condition is that $v_d\in C$. Hence, the complement is composed of only $S-1$ voters. We distinguish these coalitions by the number of active paths that they fully include, each such path adds $3n$ voters and at list one should be fully contained. Let us consider coalitions that include $k$ active paths. Then, this makes it possible for $v^*$ to be reachable by $3k$ element voters. We count the number of coalitions that include $i$ such voters (at least one must be included). The rest of voters in the coalition may be arbitrary voters in the rest of element voters, arbitrary voters in paths $\{v_{S^{t}}, t\in [3n]\}$ for $S\not\in\{T_1,\ldots,T_n\}$ or other voters in paths corresponding to sets $S\in\{T_1,\ldots,T_n\}$ given that a full path is not taken. 

\end{proof}

\section{Deferred proofs from Section~\ref{section : VWM}} \label{app : hardness2}

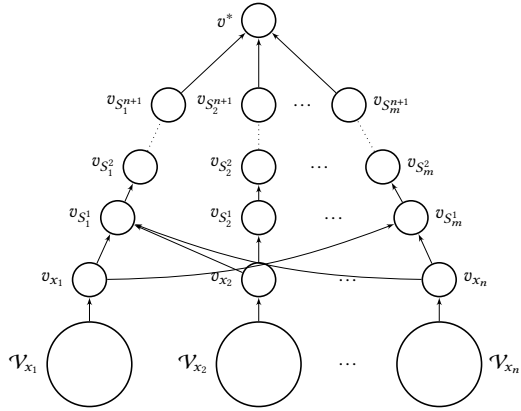
\begin{figure}[t]
\centering
\scalebox{0.75}{
\begin{tikzpicture}
\tikzset{vertex/.style = {shape=circle,draw = black,thick,fill = white,minimum size = 4mm}}
\tikzset{edge/.style = {->,> = latex'}}

\node[vertex][minimum size=.6cm, label=left:{$v^*$}] (u) at  (1,2.8) {};
\node[vertex][minimum size=.6cm, label=left:{$v_{S_{1}^{n+1}}$}] (vs1N) at  (-.6,1.3) {};
\node[vertex][minimum size=.6cm, label=left:{$v_{S_{1}^2}$}] (vs12) at  (-1.1,.2) {};
\node[vertex][minimum size=.6cm, label=left:{$v_{S_{1}^1}$}] (vs11) at  (-1.5,-.7) {};

\node[vertex][minimum size=.6cm, label=left:{$v_{S_{2}^{n+1}}$}] (vs2N) at  (1,1.3) {};
\node[vertex][minimum size=.6cm, label=left:{$v_{S_{2}^2}$}] (vs22) at  (1,.2) {};
\node[vertex][minimum size=.6cm, label=left:{$v_{S_{2}^1}$}] (vs21) at  (1,-.7) {};

\node[vertex][minimum size=.6cm, label=right:{$v_{S_{m}^{n+1}}$}] (vsmN) at  (2.6,1.3) {};
\node[vertex][minimum size=.6cm, label=right:{$v_{S_{m}^2}$}] (vsm2) at  (3.2,.2) {};
\node[vertex][minimum size=.6cm, label=right:{$v_{S_{m}^1}$}] (vsm1) at  (3.7,-.7) {};

%\node[vertex][minimum size=.6cm] (vs2) at  (1,0) {$v_{S_2}$};
%\node[vertex][minimum size=.6cm] (vs3) at  (2.7,0) {$v_{S_M}$};

%\node[vertex][minimum size=2cm] (Clique) at (5.5,0.5)  {$\mathcal{I}$};

\node[vertex][minimum size=.6cm, label=left:{$v_{x_1}$}] (v1) at  (-2,-1.8) {};
\node[vertex][minimum size=.6cm, label=left:{$v_{x_2}$}] (v2) at (1,-1.8) {};
\node[vertex][minimum size=.6cm, label=right:{$v_{x_n}$}] (vn) at (4.2,-1.8) {};

\node[vertex][minimum size=1.5cm, label=left:{$\mathcal{V}_{x_1}$}] (v21) at  (-2,-3.3) {};
\node[vertex][minimum size=1.5cm, label=left:{$\mathcal{V}_{x_2}$}] (v22) at (1,-3.3) {};
\node[vertex][minimum size=1.5cm, label=right:{$\mathcal{V}_{x_n}$}] (v2n) at (4.2,-3.3) {};

\path (vs21) to node {\dots} (vsm1);
\path (vs22) to node {\dots} (vsm2);
\path (vs2N) to node {\dots} (vsmN);
\path (v2) to node {\dots} (vn);
\path (v22) to node {\dots} (v2n);

\draw[edge] (vs1N) to (u);
\draw[edge] (vs2N) to (u);
\draw[edge] (vsmN) to (u);

\draw[edge] (v1) to (vs11);
\draw[edge] (vs11) to (vs12);
\draw[edge] (vs21) to (vs22);
\draw[edge] (vsm1) to (vsm2);
\path[draw, dotted] (vs12) edge (vs1N);
\path[draw, dotted] (vs22) edge (vs2N);
\path[draw, dotted] (vsm2) edge (vsmN);

\draw[edge] (v1)[bend right = 10] to (vsm1);
\draw[edge] (v2) to (vs11);
\draw[edge] (v2) to (vs21);
\draw[edge] (vn) [bend left = 10] to (vs11);
\draw[edge] (vn)  to (vsm1);

\draw[edge] (v21) to (v1);
\draw[edge] (v22) to (v2);
\draw[edge] (v2n) to (vn);

\end{tikzpicture}}
\caption{An example of digraph $D$ resulting from the reduction. For every $x\in U$, each set $\mathcal{V}_x$ includes $M-1$ vertices, where there exists an arc from each vertex in $\mathcal{V}_x$ to the element voter $v_x$. For each $S \in P$, there exists a directed path $(v_{S^1}, v_{S^2}, \dots, v_{S^{n+1}}, v^*)$.} \label{figPreservinghardness}
\end{figure}

\thWTPPreservingFactor*
\begin{proof}
The membership in NP is straightforward. 
Let us prove the approximation hardness result (the NP-hardness result can be proved by a similar reduction).

	    We design a reduction from the maximum coverage problem. 
	    In the maximum coverage problem, we are given a universe $ U = \{x_1,\ldots, x_n\}$ of $n$ elements, a collection $ P=\{ S_1, S_2, ..., S_m \} $ of $m$ subsets of $U$, and an integer $k$. The goal is to pick at most $k$ sets $S_{i_1},\ldots,S_{i_k}$ such that $|\bigcup_{j=1}^{k} S_{i_j}|$ is maximized. Stated otherwise, we want to cover the maximum number of elements in $U$.
	    This problem is known to be hard to approximate better than $1 - 1/e$~\cite{feige1998threshold}. 

	    \paragraph{The reduction.} 
          From an instance $I = (U,P,k)$ of the maximum coverage problem, we create the following instance $I'$ of the optimization variant of \textbf{WMaxP}. 
          As illustrated in Figure~\ref{figPreservinghardness}, the graph $D = (V,A)$ is compounded of the following elements:
          \begin{itemize}
           \item{For each element $x \in U$, we create a voter $v_{x}$. These voters will be called element voters or element vertices in the following.}
    	  \item{For each set $S \in P$, we create $(n+1)$ voters $v_{S^{1}}, \dots, v_{S^{(n+1)}}$. We create an arc from $v_{S^{i}}$ to $v_{S^{(i+1)}}$ for $i\in [n]$. These voters will be called set voters or set vertices in the following.
	      \item For every $ x \in U$ and $ S \in P $, we create an arc from $v_x$ to $v_{S^{1}}$ if  $x \in S$.}
	      \item We create a voter $v^*$. For every $ S \in P$, we create an arc from $v_{S^{(n+1)}}$ to $v^*$.
	      \item For every $x \in U$, we create a set $\mathcal{V}_x$ with $M-1$ vertices where $M:= ((n+1)m+1)T$ and $T$ is a constant the value of which will be discussed later on. We create an arc from each vertex in $\mathcal{V}_x$ to element voter $v_x$.
    	  \item We define a budget $\bar{k}=(n+1)k+n$.
     \end{itemize}		 
     All voters have weight equal to one. In the initial delegation function, we have that $\forall x \in U, d(v_x) = v_x$, $\forall S \in P,\forall i\in [n+1], d(v_{S^{i}}) = v_{S^{i}}$, and $d(v^*) = v^*$. Last, we have that $\forall x \in U$, and $\forall u\in \mathcal{V}_x, d(u) = v_x$.
	
	Let us consider a delegation function $d'$, such that $|\{i\in V:d(i)\neq d'(i)\}|\le \bar k$.
	We say that the $S$-path is active for $d'$ if $d'(v_{S^{i}}) = v_{S^{(i+1)}}$ for $i\in [n]$ and $d'(v_{S^{(n+1)}}) = v^*$. Note that the budget only allows $k$ $S$-paths to be active and we denote by $\mathcal{S}(d')$ the set of sets $S$ such that the $S$-path is active for $d'$. 
	Moreover, we will say that an element $x$ is covered by $d'$ if $d'(v_x) = v_{S^{1}}$ for some set $S\in P$ such that the $S$-path is active for $d'$. We denote by $\mathcal{U}(d')$ the set of elements covered by $d'$.
	
	It is easy to notice that $|\mathcal{U}(d')|M \le \alpha_{d'}(v^*) \le |\mathcal{U}(d')|M + M/T$ (we assume that delegations from voters in sets $\mathcal{V}_i$ do not change as this can only  be counter-productive for the problem at hand). Let us assume we have a $c$-approximation algorithm for \textbf{WMaxP}. Let $d'$ be the solution returned by this algorithm and $d^*$ be the optimal solution. Then,
	$(|\mathcal{U}(d')| + 1/T)M \ge \alpha_{d'}(v^*) \ge c \alpha_{d^*}(v^*)$.  
	
	Let $OPT$ be the optimal value of the maximum coverage instance and $X^*$ be an optimal solution. Consider the delegation function $d_{X^*}$ obtained from $d$ by making each $S$-path for which $S\in X^*$ active and making each element voters $v_x$ delegate to $v_{S^1}$ if $x$ is covered by $S$ in solution $X^*$ (the choice of $S$ can be made arbitrarily if several sets cover $x$ in $X^*$). It is clear that $d_{X^*}$ is a valid solution such that $|\mathcal{U}(d_{X^*})| = OPT$. Hence, $\alpha_{d^*}(v^*) \ge \alpha_{d_{X^*}}(v^*) \ge |\mathcal{U}(d_{X^*})|M = OPT \cdot M$. Hence, we obtain that $|\mathcal{U}(d')| + 1/T \ge c OPT$. We conclude by noticing that $\mathcal{S}(d')$ provides a solution in the maximum coverage instance covering the elements in $\mathcal{U}(d')$ and that by making $1/T$ tend towards $0$, we would obtain a $c(1-\epsilon)$ approximation algorithm for the maximum coverage problem with $\epsilon$ as close to 0 as wanted. 

\end{proof}

\WTPReqBarZero*
\begin{proof}[Proof sketch]
Let $I = (\langle D = (V,A),\omega, d , k\rangle, i^*, k, \tau)$ be an instance of \textbf{WMaxP} where $\bar{\mathtt{req}} = 0$. Note that to reach the threshold in this case, we should make all voters delegate to $i^*$ by modifying at most $k$ delegations. %Let us denote by $H_d = \{T_d(v_{1}), \ldots, T_d(v_l)\}$ the forest of initial delegations given in input, where $T_d(v_i)$ denotes a tree rooted in $v_i$ (i.e., with guru $v_i$). 

Let us consider a delegation function $d'$ such that $|\{i : d(i)\neq d'(i)\}| \le k$, and $\alpha_{d'}(i^*) = \tau$ (assuming such a solution exists). 
Even if $i^*$ is not a guru in $d$, notice that $i^*$ is necessarily a guru in $d'$. Indeed, $i^*$ needs the delegations of every other voters including the ones in $c_d(i^*, d^*_{i^*})$, and we should not have a cycle in $d'$. Hence, we can assume that $i^*$ is a guru in $d$, decreasing the budget $k$ by one if it is not the case. Now consider the weighted digraph $\bar D = (V, \bar A, \lambda)$, where 
\begin{itemize}
    \item $\bar A$ is obtained from $A$ by removing the outgoing edges from $i^*$ and then reversing all the other edges;
    \item $\lambda((u,v)) = n + 1 $ if $d(v) = u$ and $\lambda((u,v)) = n$, otherwise.  
\end{itemize}
A branching $b$ of $\bar D$ is a set of edges such that (i) if $(x_1,y_1)$, $(x_2,y_2)$ are distinct edges of $b$ then $y_1 \neq y_2$ (nodes have in-degree at most one); and (ii) $b$ does not contain a cycle. An optimum branching $b$ of  $\bar D$ is one that maximizes $\sum_{e \in b} \lambda(e)$. 
We run Edmond's algorithm on $\bar D$~\cite{edmonds1967optimum} to find an optimum branching $b$. 
Let $\Gamma(b)$ be the number of edges in $b$, and $\Omega(b)$ be the number of edges $(u,v)$ in $b$ such that $d(v) = u$. 
It is clear that $\sum_{e \in b} \lambda(e) = \Gamma(b) n + \Omega(b)$. 
As $\Omega(b) \le n-1$, an optimum branching lexicographically maximizes the value of $\Gamma$ and then the one of $\Omega$. We claim that $I$ is a yes instance iff $\Gamma(b) = n-1$, and  $\Omega(b) \ge n-1-k$.  Indeed, notice that one obtains a one to one correspondence between branchings and delegation graphs by simply reversing the arcs. Having $\Gamma(b) = n-1$, implies that there is only one root in the branching, i.e., only one vertex with no incoming edge, and hence only one guru in the corresponding delegation function. As we have removed possible delegations of $i^*$, this guru is necessarily $i^*$. Additionally, $\Omega(b) \ge n-1-k$ implies that at most $k$ delegations choices are changed w.r.t. $d$. 
\end{proof}

%\WMaxPIsWHardReqBar*

\WMaxPFPT*
\begin{proof}
Let $I = (\langle D = (V,A),\omega, d\rangle, i^*, k, \tau)$ be an instance of \textbf{WMaxP}. 
As voters' weights are positive integers, the maximum number of additional voters that $i^*$ needs the support of to reach the threshold is bounded by $\mathtt{req}$. 
We first note that one can collapse the tree $T_d(i^*)$ in one vertex with weight $\alpha_d(i^*)$. 
Moreover, note that if $i^*$ is not a guru, we then decreases $k$ by one and remove her delegation: she should be a guru if we want her accumulated weight to be different from 0. 
Let us consider a delegation function $d'$ such that $|\{i : d(i)\neq d'(i)\}| \le k$, and $\alpha_{d'}(i^*) \ge \tau$ (assuming such a solution exists). 
A subtree of $T_{d'}(i^*)$ rooted in $i^*$ with at most $\mathtt{req}+1$ voters accumulates a voting weight greater than or equal to $\tau$. We denote this tree simply by $T$. 
One can guess the shape of $T$ and  then look for this tree in $D$. Note that $T$ should be a directed tree with all vertices delegating directly or indirectly to $i^*$. 
This can be done in FPT time by adapting a color coding technique~\cite{alon1995color}. 
The idea is to color the graph randomly with $\mathtt{req}+1$ colors. 
If the tree that we are looking for is present in graph $D$, it will be colored with the $\mathtt{req}+1$ colors (i.e., one color per vertex) with some probability only dependent of $\mathtt{req}$. We say that such a tree is colorful. Let $r$ be an arbitrary vertex in $T$. 
In $O(2^{O(\mathtt{req})}k^2|A|)$ time, we find, for each vertex $v\in V$, all color sets that appear on colorful copies of $T$ in $G$ in which $v$ plays the role of $r$. 
More precisely, for each colorset $C$, we do not only keep one element but rather at most $k+1$ elements. Indeed, two trees $T_C$ and $T_C'$ rooted in $r$ with colors $C$ are not equivalent in the sense that one may have a higher accumulated weight than the other; we use $\alpha(T_C)$ to denote this weight in tree $T_C$. Moreover $T_C$ may only be part of a valid solution for \textbf{WMaxP} if $|(u,v) \in T_C, d(u) \neq v| \le k$;  we use $\delta(T_C)$ to denote $|(u,v) \in T_C, d(u) \neq v|$. Hence, for each color set $C$, we keep at most $k+1$ elements: For each $i\in[k]_0$, we keep the tree $T_C$ with highest value $\alpha(T_C)$ for $\delta(T_C) = i$. 
This is done by dynamic programming. If $T$ contains only a single vertex, this is trivial.  Otherwise, let $e = (r,r')$ be an arc in $T$.  The removal of $e$ from $T$ breaks $T$ into two subtrees $T'$ and $T''$. We recursively find, for each vertex $v \in V$, the color sets that appear on colorful copies of $T'$ in which $v$ plays the role of $r$, and the color sets that appear on colorful copies of $T''$ in which $v$ plays the role of $r'$. 
For every arc $e = (u,v)\in A$, if $C'$ is a color set that appears in $u$'s collection corresponding to $T'$ and $(\alpha(T_{C'}), \delta(T_{C'}))$ is one of the $k+1$ elements recorded for $C'$, if $C''$ is a color set that appears in $v$'s collection (corresponding to $T''$), and $(\alpha(T_{C''}), \delta(T_{C''}))$ is one of the $k+1$ elements recorded for $C''$. If $C'\cap C'' = \emptyset$ and $\delta(T_{C'})+ \delta(T_{C''}) + 1_{d(u) = v} \le k$, then $C'\cup C''$ is added to the collection of $u$ (corresponding to $T$) with pair $(\alpha(T_{C'}) + \alpha(T_{C''}), \delta(T_{C'}) + \delta(T_{C''}) + 1_{d(u) \neq v})$. Note that if an element already exists for $\delta(T_{C'}) + \delta(T_{C''}) + 1_{d(u) \neq v}$ and color set $C'\cup C''$, then this one only replaces it if it has a higher accumulated weight.  The algorithm succeeds with guess $T$ if one obtains in $i^*$ the complete set $C$ of $\mathtt{req}$ colors with an element $T_C$ with $\alpha(T_{C})\ge \tau$. 
This algorithm can then be derandomized using families of perfect hash functions~\cite{alon1995color,schmidt1990spatial}.

\end{proof}

\end{document}